\documentclass[twocolumn,amsmath,amssymb,prb]{revtex4}
\pdfoutput=1
\usepackage{graphicx}   % include figure files
\usepackage{dcolumn}    % align numerical table columns on the decimal point
\usepackage{bm}         % bold math symbols
\usepackage{rotating}

\newcommand{\cpl}{Chem.\ Phys.\ Lett.\ }

\renewcommand{\jcp}{J.\ Chem.\ Phys.\ }

\newcommand{\jpc}{J.\ Phys.\ Chem.\ }
\newcommand{\jpca}{J.\ Phys.\ Chem.\ A }
\newcommand{\jpcc}{J.\ Phys.\ Chem.\ A }

\renewcommand{\prl}{Phys.\ Rev.\ Lett.\ }

\newcommand{\jpcl}{J.\ Phys.\ Chem.\ Lett.\ }

\newcommand{\pnas}{Proc.\ Natl.\ Acad.\ Sci.\ }
\newcommand{\eqn}[1]{Eq.~(\ref{#1})}
\newcommand{\Eqn}[1]{Equation~(\ref{#1})}
\newcommand{\eqnn}[2]{Eqs.~(\ref{#1}) and (\ref{#2})}

\newcommand{\tp}[1]{$\times 10^{#1}$}

\newcommand{\bra}[1]{\big< \, #1 \, \big|}
\newcommand{\ket}[1]{\big| \, #1 \, \big>}
\newcommand{\expect}[1]{\big< \, #1 \, \big>}
\newcommand{\shortt}{$t \to 0_+$\ }
\newcommand{\shortta}{$t \to 0_+$}

\newcommand{\cfs}{C_{\rm fs}(t)}
\newcommand{\kQ}{k_{\rm Q}(\beta)}
\newcommand{\longt}{$t \to \infty$\ }
\newcommand{\fq}{$f({\bf q})$\ }
\newcommand{\largen}{$N \to \infty$\ }

\begin{document} 

\title{Derivation of a true ($t\rightarrow 0_+$) quantum transition-state theory. Part I: Uniqueness and equivalence to ring-polymer molecular dynamics transition-state-theory}
\author{Timothy J.~H.~Hele and Stuart C.~Althorpe\footnote{Corresponding author: sca10@cam.ac.uk}}
\affiliation{Department of Chemistry, University of Cambridge, Lensfield Road, Cambridge, CB2 1EW, UK.}
\date{\today}

\begin{abstract}
Surprisingly, there exists a  quantum flux-side time-correlation function which has a non-zero $t\rightarrow 0_+$ limit, and thus yields a rigorous quantum generalization of classical  transition-state theory (TST). In this Part I of two articles, we introduce the new time-correlation function, and derive its $t\rightarrow 0_+$ limit. The new ingredient is a generalized Kubo transform which allows the flux and side dividing surfaces to be  the 
same function of path-integral space. Choosing this common dividing surface to be a single point gives a $t\rightarrow 0_+$ limit which is identical to an expression introduced on heuristic grounds  by Wigner in 1932, but which does not give positive-definite quantum statistics, causing it to fail while still in the shallow-tunnelling regime.  Choosing the dividing surface to be invariant to imaginary-time translation gives, uniquely, a $t\rightarrow 0_+$ limit that gives the correct positive-definite quantum statistics at all temperatures, and which is {\em identical} to ring-polymer molecular dynamics (RPMD) TST. We find that the RPMD-TST rate is not a strict upper bound to the exact quantum rate, but a good approximation to one if real-time coherence effects are small. Part II will show that the RPMD-TST rate  is equal to the exact quantum rate in the absence of recrossing. \emph{Copyright (2013) American Institute of Physics. This article may be downloaded for personal use only. Any other use requires prior permission of the author and the American Institute of Physics. The following article appeared in the Jounral of Chemical Physics, 138, 084108 (2013) and may be found at http://link.aip.org/link/?JCP/138/084108/1}
\end{abstract}

\maketitle

\section{Introduction}
\label{sec:intro}

It is well known \cite{green,daan} that classical transition state theory\cite{math}  (TST) corresponds to
 taking the short time ($t\rightarrow 0_+$) limit of the classical flux-side time-correlation function, and that this 
 can be done because this function is odd and discontinuous about $t=0$ (see Fig.~1).

\newlength{\figwidths}
\setlength{\figwidths}{0.45\columnwidth}
\begin{figure}[tb]
 \resizebox{\figwidths}{!} {\includegraphics[angle=270]{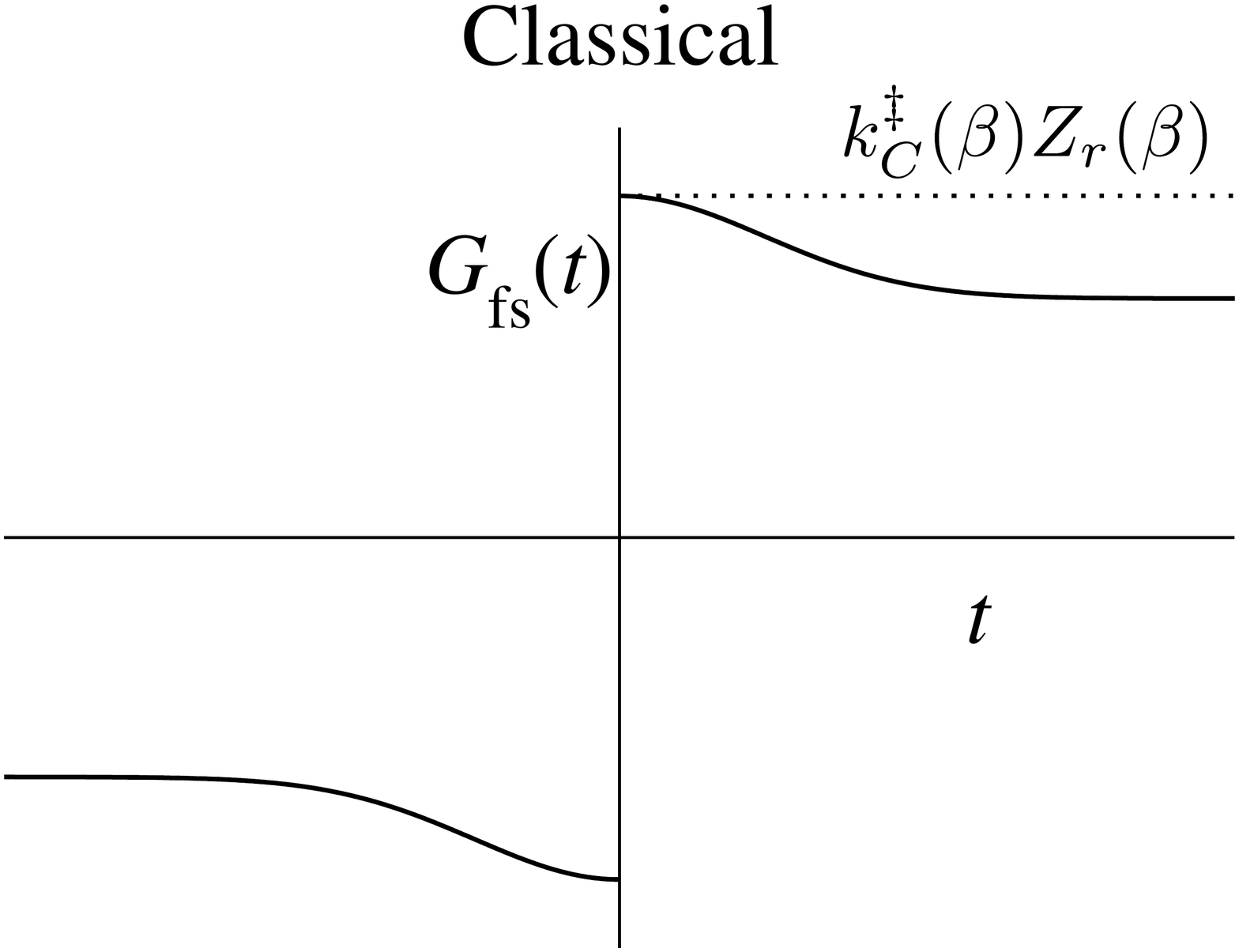}}
 \resizebox{\figwidths}{!} {\includegraphics[angle=270]{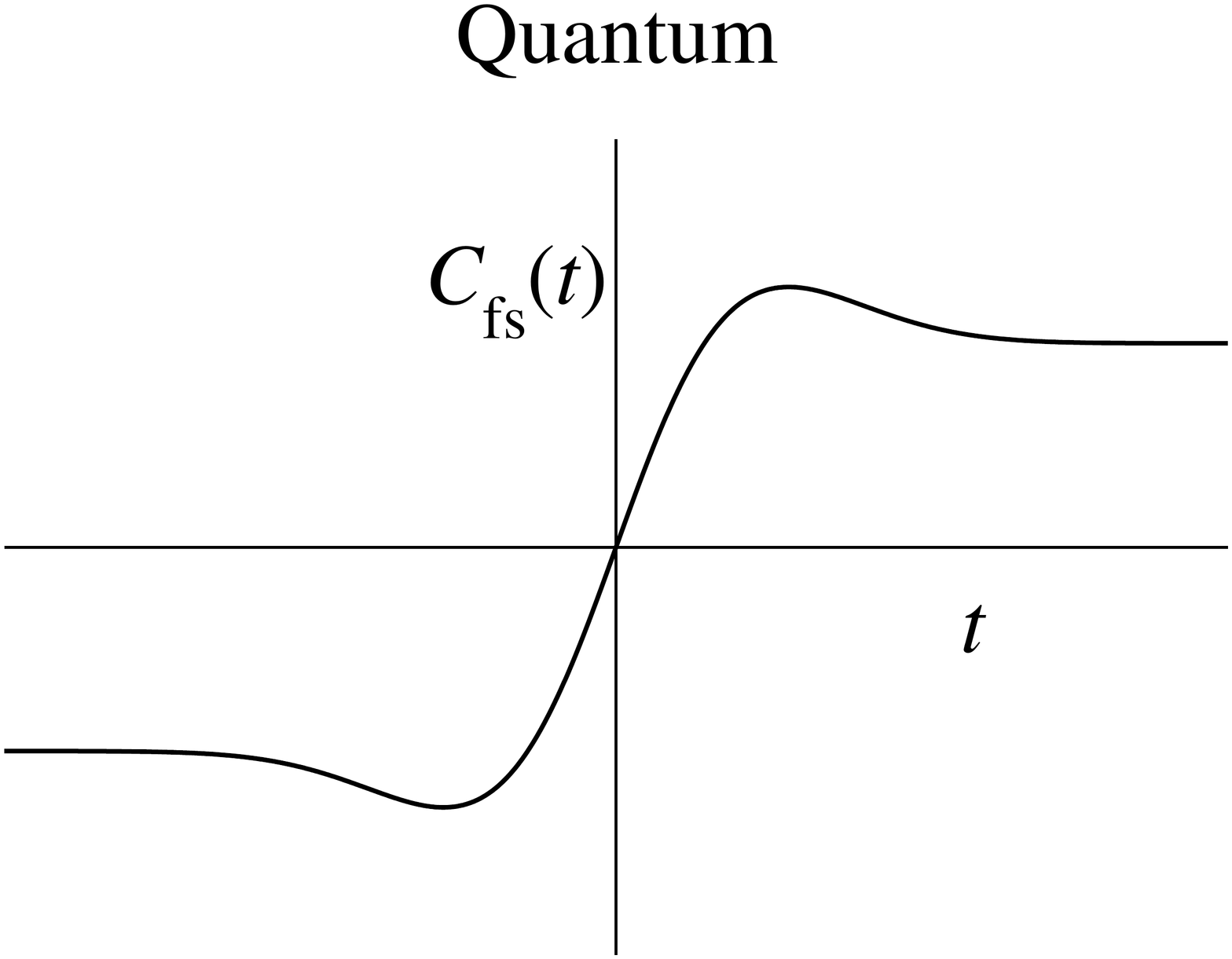}}
 \caption{Behaviour of classical and quantum flux-side time-correlation functions $G_{\rm fs}(t)$ and $C_{\rm fs}(t)$ as a function of  time $t$. $k_{\rm C}^\ddag(\beta)$ is the classical TST rate and $Z_{\rm r}(\beta)$ the reactant partition function.}
\end{figure}
 
Unfortunately, this convenient behaviour has not seemed to carry over into quantum rate-theory, \cite{MST,mill,pollack} where the various forms of quantum flux-side time-correlation function tend smoothly to zero in the $t\rightarrow 0_+$ limit (see Fig.\ 1). This is  a pity,  because it is still impossible to evaluate the
quantum flux-side time-correlation function at $t>0$ for all but the simplest systems.\cite{rates1,makri,bow,manthe} A vast number of methods have been developed in an attempt to circumvent this problem. One type of method is to 
 include quantum corrections into classical TST, \cite{truh1, truh2}  the most famous instance of this being the Wigner-Eyring formula. \cite{beloved} A more systematic approach is to use semi-classical theories \cite{billact, billhandy,stanton,bill,cole,benders,jor,jonss,spanish,kastner1,kastner2,equiv,scivr1,scivr2,scivr3} (some of which we discuss below), or the quantum instanton approach. \cite{QI1,QI2} However, there is one approach \cite{jor,rates,refined,azzouz,bimolec,ch4,mustuff,anrev,yury,tommy1,tommy2,
 stecher,gillan1,gillan2,centroid1,centroid2,schwieters}
 that is especially relevant to this article, which is to note that  classical TST  takes the form of a free-energy combined with a flux factor, and then to exploit the fact that it is relatively easy to compute a quantum free energy, using `ring-polymer' path-integration. \cite{chandler,parrinello,ceprev,charu}

This last approach has turned out to be extremely powerful. Its most general formulation is ring-polymer molecular dynamics (RPMD) TST, \cite{jor,rates,refined,azzouz,bimolec,ch4,mustuff,anrev,yury,tommy1,tommy2,stecher} in which the free-energy is obtained by constraining the ring-polymers using a general dividing surface. For reasonably symmetric reaction barriers, this dividing surface can be made a function of
 just the polymer-bead centroids. \cite{gillan1,gillan2,centroid1,centroid2,schwieters} This `centroid-TST' approach was
 the earliest form of the theory, but it breaks down for asymmetric reaction barriers.  The more general RPMD-TST approach is usually implemented indirectly,  \cite{rates,refined,azzouz,bimolec,ch4,mustuff,anrev,yury,tommy1,tommy2,stecher} as it is more convenient to simulate the system as a classical rate-process taking place in a fictitious, extended, space of ring-polymers, and to exploit the property that the resulting rate is an approximation to the RPMD-TST rate. \cite{ie} This approach has recently allowed approximate quantum rates to be computed for hydride-transfer in enzymes \cite{tommy1} and for condensed-phase electron-transfer. \cite{tommy2}

To date, however, there has been no first-principles derivation of the RPMD-TST rate,  beyond proving that it interpolates between various limits. It gives the exact classical and parabolic-barrier rates at high temperatures, \cite{rates} and is close to the `Im F' instanton rate \cite{cole,jor,benders,equiv,mills} at low temperatures.   Note that the `Im F' instanton rate has itself not been derived from quantum rate-theory; it gives good approximations to the rate in the deep-tunnelling regime, and is probably best regarded as a rough interpolation between Miller's steepest-descent instanton theory\cite{bill} and classical rate theory.

The aim of this article (Part I) and of the forthcoming Part II is to show that, contrary to what has just been said, there does exist a form of quantum flux-side time-correlation function with a non-zero  \shortt limit, which limit does give the exact rate in the absence of recrossing, and which, remarkably, is  {\em identical} to RPMD-TST. This article will focus on deriving the \shortt limit and establishing that it is equal to RPMD-TST. Part II will prove that the RPMD-TST rate is equal to the exact quantum rate in the absence of recrossing.

To obtain these new results, we will use the same overall strategy that Miller and co-workers\cite{MST,mill} used to first obtain quantum rate-theory; i.e.\ we will derive the theory for quantum scattering systems, because rate-theory is then
formally exact (since the flux-side time-correlation function has a plateau extending to $t\rightarrow\infty$), and hence straightforward to derive. We will then assume that the theory also applies to condensed-phase rate-processes, subject to the usual caveat of there being a  separation in timescales between barrier crossing and equilibration.\cite{isomer} Hence although we will make use of quantum scattering theory as a derivational tool, the quantum TST results that we obtain from it are applicable generally. Note that this article is almost completely free of quantum scattering theory, but it will appear in quite some detail in Part II.%One could almost certainly rederive the theory given here and Parts II and III using linear response theory, and in fact the main result of this article is a flux-side time-correlation function which contains a generalization of the Kubo transform.

An important theme running through this article is that of {\em linearization}, by which we mean use of the linearized semi-classical initial-value representation. \cite{scivr1,scivr2,scivr3,geva} In this approach, Feynman paths in the forward and backward real-time propagators are assumed to cancel out unless they are very close together, such that the difference in their actions can be expanded to linear order in position. This approximation has the effect of removing all real-time Feynman paths, except the forward-backward pairs that lie along classical trajectories joining the start and end points, which reduces the time-correlation function to its classical Wigner approximation. Now, as $t\rightarrow 0_+$, linearization becomes {\em exact}, meaning that the time-evolution of
the quantum flux-side time-correlation function becomes entirely classical in this limit (whereas the statistics, of course, remain quantal). We will show that this property gives rise to a \shortt quantum TST.

The article is structured as follows: After summarising some basic aspects of classical TST in Sec.\ II, we describe in Sec.\ III how the exact linearization of the dynamics in the \shortt limit
  can be used to explain why the standard flux-side time-correlation function
   tends smoothly to zero as $t\to 0_+$, and how one can modify it to make it discontinuous instead. This gives rise to a quantum TST which is identical to an expression introduced on heuristic grounds by Wigner, \cite{wiggie} but which gives bad quantum statistics at low temperatures. In Sec.~\ref{sec:QTSTstat}, we show that the correct (i.e.\ positive-definite) quantum statistics can be obtained using a generalized Kubo-transform, which gives a flux-side time-correlation function that is invariant to imaginary-time translation; the \shortt limit of this expression is identical to RPMD-TST. In Sec.~V we point out that the results of Secs.~III and IV, which were derived in one-dimension to simplify the algebra, can be generalized immediately to multi-dimensions. Section VI concludes the article.

\section{Classical transition-state theory}
\label{sec:clastst}

Here, we discuss some  well-known ideas from classical rate theory, which we will apply in Secs \ref{sec:wigner} and \ref{sec:QTSTstat} to quantum rate theory (where their use is less intuitive). Note that this Section considers an $f$-dimensional classical system, whereas Secs.~\ref{sec:wigner}--\ref{sec:QTSTstat} consider a 1-dimensional quantum system. This is because the \shortt  limit of the latter is similar to the \shortt limit of a multi-dimensional classical system.  (Extension of the quantum theory to multi-dimensions is straightforward and is left until Sec.~\ref{sec:multid}).

The exact classical rate coefficient $k_{\rm C}(\beta)$ is given by\cite{green,daan}
\begin{align}
k_{\rm C}(\beta)Z_{\rm r}(\beta)&=\lim_{t\rightarrow\infty}-{d\over dt}G_{\rm ss}(t)
\nonumber\\&=\lim_{t\rightarrow\infty}G_{\rm fs}(t)
\end{align}
where $Z_{\rm r}(\beta)$ is the (classical) reactant partition function, $\beta=1/k_{\rm B}T$, $G_{\rm ss}(t)$ is the side-side time-correlation function
\begin{align}
G_{\rm ss}(t)={1\over(2\pi\hbar)^f}\int \!{d{\bf p}}\int \!{d{\bf q}}\,e^{-\beta H}h[s({\bf q})]h[s({\bf q}_t)]
\end{align}
and $G_{\rm fs}(t)$ is the flux-side time-correlation function
\begin{align}
G_{\rm fs}(t)={1\over(2\pi\hbar)^f}\int \!{d{\bf p}}\int \!{d{\bf q}}\,e^{-\beta H}\delta[s({\bf q})]
{\dot s({\bf q},{\bf p})}h[s({\bf q}_t)]
\end{align}
The $t=0$ time-derivative of $s({\bf q})$ is defined to be
\begin{align}
{\dot s}({\bf q},{\bf p})=\sum_{i=1}^f {p_i\over m_i}{\partial s({\bf q})\over\partial q_i}
\end{align}
and is the flux through the dividing surface at time ${t\rightarrow 0_+}$; ${\bf q}_t$ is the position at time $t$ of a classical particle that started at $({\bf q,p})$ at time $t=0$.

It is well known that $G_{\rm fs}(t)$ is discontinuous in the \shortt limit, behaving as in Fig.~1, and that the classical TST rate coefficient $k_{\rm C}^\ddag(\beta)$ is given by 
\begin{align}
k_{\rm C}^\ddag(\beta)Z_{\rm r}(\beta)&=\lim_{t\rightarrow 0_+}G_{\rm fs}(t)
\end{align}
In the absence of recrossing, $G_{\rm fs}(t)$ would remain constant for all time, and thus $k_{\rm C}^\ddag(\beta)$ would be the exact rate. In a real system, for which there is recrossing, $k_{\rm C}^\ddag(\beta)$ is a strict upper bound, since recrossing of the dividing surface necessarily reduces the rate.

Now, it is useful to clarify why $G_{\rm fs}(t)$ is discontinuous at $t=0$, since this mathematical oddity is essential to the existence of classical TST and is the thing that we need to introduce into quantum  rate theory if we are to obtain a true quantum TST. To take the \shortt limit of $G_{\rm fs}(t)$, we note that
\begin{align}
\lim_{t\rightarrow 0_+} s({\bf q}_t)={\dot s}({\bf q},{\bf p})\, t+s({\bf q})
\end{align}
and hence that
\begin{align}
\lim_{t\rightarrow 0_+}\delta[s({\bf q})]h[s({\bf q}_t)]&=\delta[s({\bf q})]\,h\!\left[{\dot s}({\bf q},{\bf p})\, t+s({\bf q})\right]\nonumber\\&=
\delta[s({\bf q})]\,h[{\dot s({\bf q},{\bf p})}]
\end{align}
which gives the familiar expression
\begin{align}
\lim_{t\rightarrow 0_+}G_{\rm fs}(t)=&{1\over(2\pi\hbar)^f}\int \!{d{\bf p}}\int \!{d{\bf q}}\nonumber \\
& \times \,e^{-\beta H}\delta[s({\bf q})]
{\dot s({\bf q},{\bf p})}h[{\dot s({\bf q},{\bf p})}]
\end{align}
Hence the discontinuity at \shortt arises because the Dirac-delta function sets $s({\bf q})$ to zero inside the Heaviside function, turning the latter into a time-independent momentum [i.e.\ ${\dot s({\bf q},{\bf p})}$] filter, which is abruptly switched off at $t=0$. Of course the reason this happens is that the flux and side dividing surfaces are the {\em same function} $s({\bf q})$. If we choose different functions $s_1({\bf q})$ and $s_2({\bf q})$ for the dividing surfaces, we obtain
\begin{align}
\lim_{t\rightarrow 0_+}\delta[s_1({\bf q})]h[s_2({\bf q}_t)]&=\delta[s_1({\bf q})]\,h\!\left[{\dot s_2}({\bf q},{\bf p})\, t+s_2({\bf q})\right]\nonumber\\&\ne
\delta[s_1({\bf q})]\,h[{\dot s_2({\bf q},{\bf p})}]
\end{align}
i.e.\ the momentum filter  switches off smoothly as \shortt (because the contribution of ${\dot s({\bf q},{\bf p})}$ inside the heaviside function is gradually turned off as $t\rightarrow 0_+$).

The property that $s_1({\bf q})$ and $s_2({\bf q})$ must be the same to give a non-zero \shortt limit is not restricted to the Boltzmann distribution; any (non-pathological) distribution of $({\bf q},{\bf p})$ will also behave this way. As a result, we can use the analysis just given (which is trivial in the context of classical rate theory) to modify the quantum flux-side
time-correlation function so that it also is non-zero in the \shortt limit.

\section{Wigner-Miller transition-state theory}
\label{sec:wigner}

\subsection{Exact quantum rate theory}
\label{ssec:exactqm}

To simplify the algebra, we consider a 1-dimensional quantum scattering system with mass $m$,  hamiltonian $\hat H$, and a potential $V(q)$ which tends asymptotically to fixed limits, and has a barrier. However, we emphasise that everything  we derive below generalises immediately to multi-dimensional systems (see Sec.~\ref{sec:multid}), and also applies to condensed-phase systems (see the comments in the Introduction).

The exact quantum rate coefficient $\kQ$ is given by\cite{MST}
\begin{align}\label{baby}
k_{\rm Q}(\beta)Q_{\rm r}(\beta)&=\lim_{t\rightarrow\infty}-{d\over dt}C_{\rm ss}(t)
\nonumber\\&=\lim_{t\rightarrow\infty}C_{\rm fs}(t)
\end{align}
where the quantum side-side and flux-side time-correlation functions are 
\begin{align}
C_{\rm ss}(t)={\rm Tr}\left[e^{-\beta{\hat H}/2}{\hat h}(q^\ddag)e^{-\beta{\hat H}/2}e^{i{\hat H}t/\hbar}{\hat h}(q^\ddag)e^{-i{\hat H}t/\hbar}\right]
\end{align}
and 
\begin{align}
C_{\rm fs}(t)={\rm Tr}\left[e^{-\beta{\hat H}/2}{{\hat F(q^\ddag)}}e^{-\beta{\hat H}/2}e^{i{\hat H}t/\hbar}{\hat h}(q^\ddag)e^{-i{\hat H}t/\hbar}\right]
\label{eq:cfsmil} % Miller's flux-side correlation function
\end{align}
and where\begin{align}
{\hat F(q^\ddag)}={1\over 2 m}\left[{\hat p}\,\delta({\hat q}-q^{\ddag})+\delta({\hat q}-q^{\ddag})\,{\hat p}  \right]
\end{align}
is the flux operator, and ${\hat h}(q^{\ddag}) =  h({\hat q}-q^{\ddag})$ the side operator.

Now let us consider the \shortt limit. We rewrite $\cfs$ as
\begin{align}
C_{\rm fs}(t)=&\int_{-\infty}^{\infty}\! d{q_2}\, \int_{-\infty}^{\infty}\! d{z_2}\,\int_{-\infty}^{\infty}\! d{\Delta_2}\, h(z_2-q^\ddag)\nonumber\\
&\times \expect{q_2-\Delta_2/2|e^{-\beta{\hat H}/2}{\hat F(q^\ddag)}e^{-\beta{\hat H}/2}|q_2+\Delta_2/2}\nonumber\\
&\times \expect{q_2+\Delta_2/2|e^{i{\hat H}t/\hbar}|z_2} \nonumber \\
&\times \expect{z_2|e^{-i{\hat H}t/\hbar}|q_2-\Delta_2/2} 
\label{eq:cfspi} % cfs path integral
\end{align}
(where the reason for the subscripts will become clear shortly).

We then use the relations
\begin{align}
\lim_{t\rightarrow 0_+}e^{-i{\hat H}t/\hbar}=e^{-i{\hat K}t/\hbar}e^{-i{\hat V}t/\hbar}
\label{eq:kv} % K and V operators
\end{align}
where $\hat V$ and $\hat K$ are the potential and kinetic energy operators respectively, and
\begin{align}
\expect{x|e^{-i{\hat K}t/\hbar}|y}=\sqrt{m\over 2\pi i\hbar t}e^{im(x-y)^2/2\hbar t}
\label{eq:k} % Evaluation of K matrix element 
\end{align}
to obtain
\begin{align}\label{pixie}
\lim_{t\rightarrow 0_+}&C_{\rm fs}(t)={1\over 2\pi\hbar}\int_{-\infty}^{\infty}\! d{q_2}\, \int_{-\infty}^{\infty}\! d{p_2}\,\int_{-\infty}^{\infty}\! d{\Delta}_2\, \nonumber \\
& \times h(q_2+p_2t/m-q^\ddag)
e^{ip_2\Delta_2/\hbar}\nonumber\\
&\times \expect{q_2-\Delta_2/2|e^{-\beta{\hat H}/2}{\hat F(q^\ddag)}e^{-\beta{\hat H}/2}|q_2+\Delta_2/2}
\end{align}
This procedure is an example of {\em linearization}:\cite{scivr2,scivr3,geva,wiggie} we obtain a time-correlation function involving a Wigner transform of the  Boltzmannized flux operator, in which the dynamics is now classical (i.e.\ $z_2\equiv q_{2t}= q_2+p_2t/m$). At finite times linearization is approximate, but in the \shortt limit taken here it is clearly exact. 

The statistics in the \shortt limit is therefore quantum, but the dynamics is classical. However, it is difficult to use the reasoning of Sec.~\ref{sec:clastst} to explain why $\cfs$ tends smoothly to zero in the \shortt limit, since the  flux operator $\hat F$ is embedded in the quantum Boltzmann operator. Fortunately, we can easily change this by inserting  the identity 
\begin{align}
{\hat I} = & \int_{-\infty}^\infty \!dq_1\int_{-\infty}^\infty \!dz_1\int_{-\infty}^\infty \!d\Delta_1\,\nonumber \\ 
& \times \ket{q_1+\Delta_1/2} \expect{q_1+\Delta_1/2|e^{i{\hat H}t/\hbar}|z_1}\nonumber\\
& \times \expect{z_1|e^{-i{\hat H}t/\hbar}|q_1-\Delta_1/2}\bra{q_1-\Delta_1/2}
\label{eq:iden}
\end{align}
This converts \eqn{eq:cfspi} into the expression
\begin{align}\label{belly}
C_{\rm fs}(t)=\int\! d{\bf q}\, &\int\! d{\bf z}\,\int\! d{\bf \Delta}\,{\cal \hat F}(q_1-q^\ddag)h(z_2-q^\ddag)\nonumber\\ 
\times\prod_{i=1}^{2}&\expect{q_{i-1}-\Delta_{i-1}/2|e^{-\beta{\hat H}/2}|q_i+\Delta_i/2}
\nonumber\\ 
\times &\expect{q_i+\Delta_i/2|e^{i{\hat H}t/\hbar}|z_i} \nonumber \\
\times &\expect{z_i|e^{-i{\hat H}t/\hbar}|q_i-\Delta_i/2}
\end{align}
where $\int\! d{\bf q}\equiv\int_{-\infty}^\infty\! dq_1\int_{-\infty}^\infty\! dq_2$ etc., and
\begin{align}
{\cal \hat F}(q_1-q^\ddag)={1\over 2m}\left[{\hat p_1}\delta(q_1-q^\ddag)+\delta(q_1-q^\ddag){\hat p_1}\right]
\label{eq:fluxdef} % Flux Definition
\end{align}
where we introduce the convention that the first term in ${\cal \hat F}(q_1-q^\ddag)$ (after the square bracket) is inserted
 between $e^{-\beta\hat H/2}\ket{q_1+\Delta_1/2}$ and $\bra{q_1+\Delta_1/2}e^{i{\hat H}t/\hbar}$ in \eqn{belly}, with ${\hat p_1}$ acting {\em only} on $\ket{q_1+\Delta_1/2}$ (via its adjoint), and  the second term is inserted between $e^{-i{\hat H}t/\hbar}\ket{q_1-\Delta_1/2}$ and $\bra{q_1-\Delta_1/2}e^{-\beta\hat H/2}$, with ${\hat p_1}$ acting only on $\bra{q_1-\Delta_1/2}$. We will use this convention throughout the article.

\begin{figure}[tb]
 \resizebox{\columnwidth}{!} {\includegraphics[angle=270]{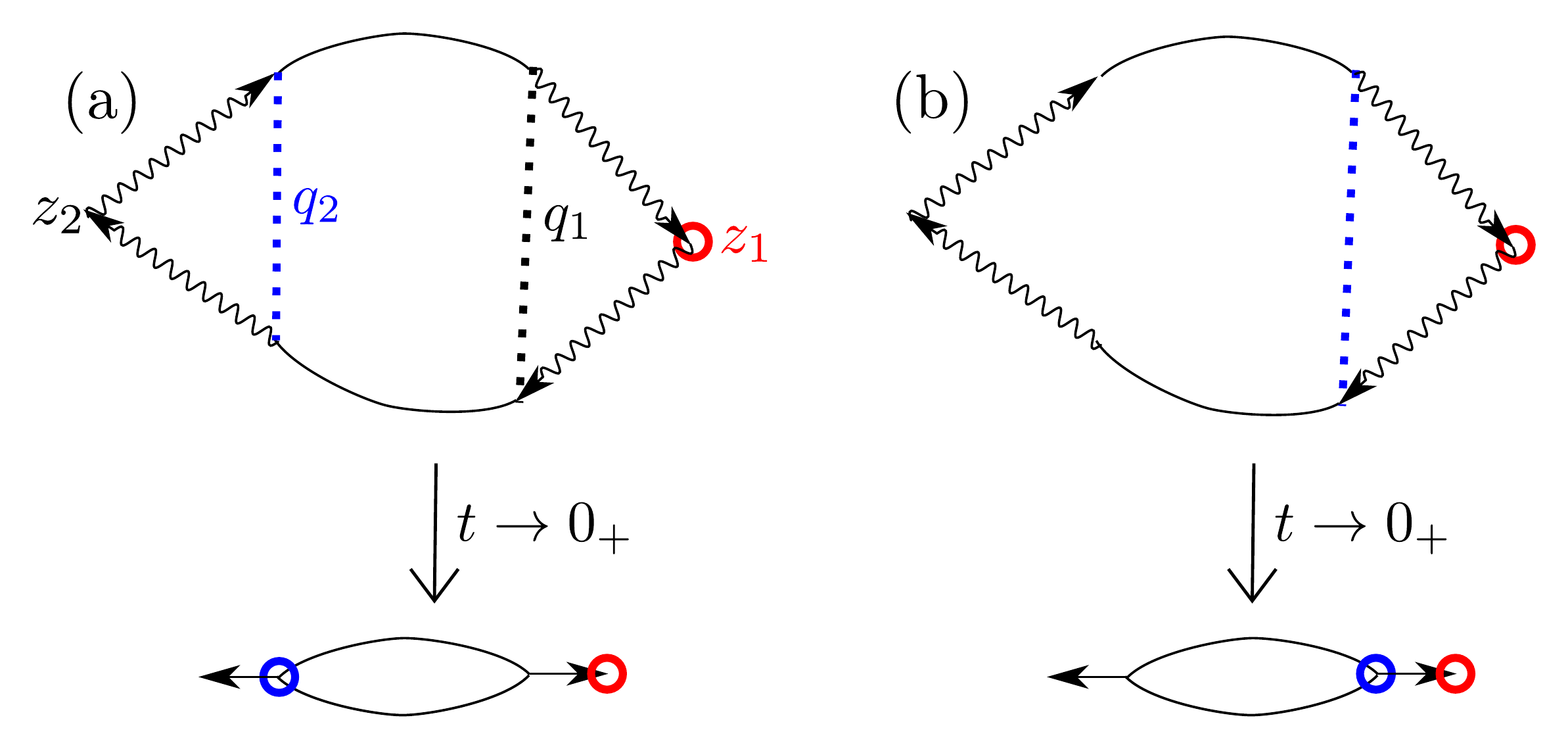}}
 \caption{Schematic representations of the quantum flux-side time-correlation functions in (a) \eqn{eq:mil2b} and (b) \eqn{eq:cfsnf}, showing the effect of making the flux and side dividing surfaces the same function of path-integral space. The solid lines 
  represent imaginary-time propagation and the wavy arrows forward-backward real-time propagation. The lower pair of diagrams represent the upper pair in the \shortt limit. In (a) the flux and side dividing surfaces (blue line and red circle)  are on opposite sides of the imaginary-time path-integral, which gives a zero \shortt limit; in (b) they are in the same place, giving a non-zero \shortt limit and hence the quantum TST of \eqn{eq:kwm}.}
\end{figure}

\Eqn{belly} is identical to \eqn{eq:cfspi}, since all we have done is insert an identity, such that the imaginary time has been split into two sections, separated by a forward-backward real-time section (see Fig.~2). 
Using the relations in \eqnn{eq:kv}{eq:k} to take the \shortt limit, we obtain 
\begin{align}
\lim_{t\rightarrow 0_+}&C_{\rm fs}(t)={1\over(2\pi\hbar)^2}\int\! d{\bf q}\, \int\! d{\bf p}\,\int\! d{\bf \Delta}\, \nonumber \\
&\times {p_1\over m}\delta(q_1-q^\ddag)h(q_2+p_2t/m-q^\ddag)\nonumber\\ 
&\times\prod_{i=1}^{2}e^{ip_i\Delta_i/\hbar}\expect{q_{i-1}-\Delta_{i-1}/2|e^{-\beta{\hat H}/2}|q_i+\Delta_i/2}
\label{eq:mil2b} % Miller's expression written with two beads
\end{align}
which is clearly identical to \eqn{pixie} [and may be obtained from it more directly by inserting momentum states next
to ${\hat F}(q^\ddag)$, but we prefer this more circuitous route, for reasons which will become clear in Sec.~\ref{ssec:movedivsurf}].

Equation \ref{eq:mil2b} shows why $\cfs$ tends smoothly to zero as $t\rightarrow0_+$: it is because $\cfs$ is equivalent to the short-time classical flux of a two dimensional system in which the flux and side dividing surfaces are {\em different} functions of the coordinates $\bf q$;
more formally, these surfaces are different functions of (imaginary-time) path-integral space.

\subsection{Derivation of Wigner-Miller TST}
\label{ssec:movedivsurf}

It follows from the above that if the flux and side dividing surfaces are the {\em same} function of path-integral space then the flux-side time-correlation function must have a discontinuity at $t=0$, and thus a non-zero \shortt limit. It is easy to modify $C_{\rm fs}(t)$ so that it has this property; one simply changes the flux dividing surface from $q_1=q^\ddag$ to $q_2=q^\ddag$ [see Fig.~2b].  The resulting flux-side time-correlation function is then
\begin{align}
C_{\rm fs}^{[1]}(t)=&\int_{-\infty}^{\infty}\! d{q}\, \int_{-\infty}^{\infty}\! d{z}\,\int_{-\infty}^{\infty}\! d{\Delta}\, {\cal \hat F}(q-q^\ddag)h(z-q^\ddag)\nonumber\\
&\times \expect{q-\Delta/2|e^{-\beta{\hat H}}|q+\Delta/2}\nonumber\\
&\times \expect{q+\Delta/2|e^{i{\hat H}t/\hbar}|z}
\expect{z|e^{-i{\hat H}t/\hbar}|q-\Delta/2}
\label{eq:cfsnf} % cfs New Form
\end{align}
with ${\cal \hat F}(q-q^\ddag)$ defined as in \eqn{eq:fluxdef} and the $[1]$-superscript indicating that the imaginary time
propagator has been split just once. It is straightforward to show (see Part II) that
\begin{align}
k_{\rm Q}(\beta)Q_{\rm r}(\beta)=\lim_{t\rightarrow\infty} C_{\rm fs}^{[1]}(t)
 \end{align}
and hence that $C_{\rm fs}^{[1]}(t)$ is a valid flux-side time-correlation function whose $t\rightarrow\infty$ limit gives the exact quantum rate. 
Since the \shortt limit of $C_{\rm fs}^{[1]}(t)$ is non-zero, we can define a quantum TST rate
\begin{align}
k^\ddag_{\rm WM}(\beta)Q_{\rm r}(\beta)=\lim_{t\rightarrow 0_+}C_{\rm fs}^{[1]}(t)
\end{align}
It is clear that $k^\ddag_{\rm WM}(\beta)$ is equal to the exact rate $k_{\rm Q}(\beta)$ if there is no recrossing of the dividing surface (since then, by definition, $C_{\rm fs}^{[1]}(t)$ is constant in time). Hence it would seem that we already have a working quantum generalization of classical TST.

However,  the explicit form of $k^\ddag_{\rm WM}(\beta)$ [obtained by applying the relations in \eqnn{eq:kv}{eq:k} to \eqn{eq:cfsnf}] turns out to be
\begin{align}
k^\ddag_{\rm WM}(\beta)Q_{\rm r}(\beta)=&{1\over 2\pi\hbar}\int_{-\infty}^{\infty}\! d{q}\, \int_{-\infty}^{\infty}\! d{p}\,\int_{-\infty}^{\infty}\! d{\Delta}\,\nonumber \\
& \times {p\over m}\delta(q-q^\ddag) h(p) e^{ip\Delta/\hbar}\nonumber\\
&\times \expect{q-\Delta/2|e^{-\beta{\hat H}}|q+\Delta/2}
\label{eq:kwm} % k Wigner Miller
\end{align}
This is a well-known expression, having been introduced by Wigner\cite{wiggie,WM} in 1932, and proposed later by Miller\cite{mill} as a possible quantum transition-state generalization to TST.  Neither of these authors was aware of \eqn{eq:cfsnf} which is new (to the best of our knowledge). Instead,  they arrived at \eqn{eq:kwm} using heuristic arguments, based on quantum-classical correspondence. We therefore expect $k^\ddag_{\rm WM}(\beta)$ to give a good approximation to $k_{\rm Q}(\beta)$ only at high temperature, and this turns out to be the case.

 Figure~3a plots $C_{\rm fs}^{[1]}(t)/Q_{\rm r}(\beta)$ for the example of an Eckart barrier [with $V(q)=V_0\, \text{sech}^2(q/a)$,  $V_0=0.425$ eV, $a = 0.734$ $a_0$, and  $m=1061$ $m_e$], at an inverse temperature $k_{\rm B}\beta = 1\times 10^{-3} $K$^{-1}$.  Clearly  the \shortt limit
$k^\ddag_{\rm WM}(\beta)$ is an excellent approximation to the exact rate at this (high) temperature. However, at a lower temperature of $k_{\rm B}\beta = 3 \times 10^{-3}$K$^{-1}$ (Fig.~3b), $k^\ddag_{\rm WM}(\beta)$ has already completely broken down, giving a negative estimate of the rate.
 We will show below that \eqn{eq:kwm} is only guaranteed to work above $2T_{\rm c}$, where $T_{\rm c}$ is the cross-over temperature (see Sec.~\ref{ssec:qmstat}).

 \begin{figure}[tb]
 \resizebox{0.6\columnwidth}{!} {\includegraphics{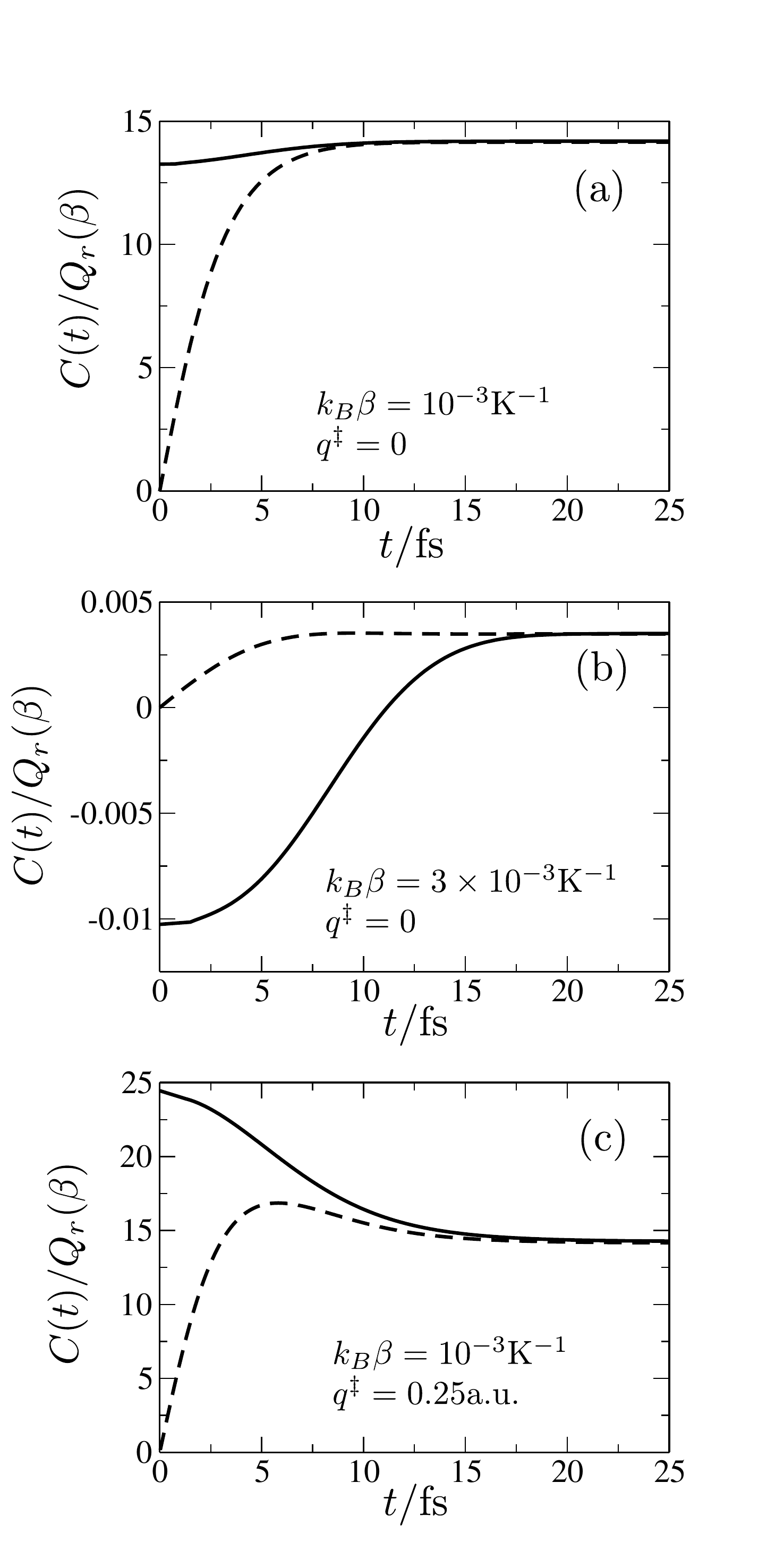}}
 % Note that, for the bottom two graphs, the $t\to0_+$ Wigner data has been spliced with the real-time calculation
 % data from tjhh2@espeto:~/Projects/NewForm/polished/graphplay/rate2.dat
 % data from /home/tjhh2/Projects/1.5form/lowt/graphplay/milnf1.5.g
 % data from tjhh2@espeto:~/Projects/NewForm/moveqdd/archive/rate.dat5
 \caption{Plots of $C(t) /Q_{\rm r}(\beta)$ versus time, where $C(t)$ represents either the flux-side time-correlation function $C_{\rm fs}^{[1]}(t)$ of \eqn{eq:cfsnf} (solid black line), or $C_{\rm fs}(t)$ of \eqn{eq:cfsmil} (dashed line). The correlation functions were computed for an Eckart barrier (see text), at the inverse-temperatures $\beta$ and dividing surface-positions $q^\ddag$ shown.}
\end{figure}
 
Before discussing this breakdown, however, we pause to review the properties of \eqn{eq:kwm} in the temperature range ($T > 2T_{\rm c}$) in which it does work, since these are the properties of a well-behaved \shortt quantum TST, and are shared by the more general \shortt quantum TST that we derive in Sec.~\ref{sec:QTSTstat}.

First, $C_{\rm fs}^{[1]}(t)$ is time-independent for a free-particle, for which \eqn{eq:kwm} thus gives the exact rate. The same is true for a parabolic barrier, provided the dividing surface is chosen such that $q^\ddag=0$ (and also that $T > 2T_{\rm c}$ --- we return to the case of $T < 2T_{\rm c}$ below). It is straightforward (though algebraically messy) to prove this last result: one uses the properties that \eqn{eq:kwm} is exact for a parabolic barrier\cite{wiggie} (with $q^\ddag=0$ and  $T > 2T_{\rm c}$), and that linearization gives the exact quantum dynamics  for a parabolic barrier;\cite{scivr1} it then follows that \eqn{eq:cfsnf} is time-independent, since a classical trajectory originating at $q^\ddag=0$ cannot recross the dividing surface. 

Second, for any realistic system, \eqn{eq:cfsnf} will not be time-independent because there will be some recrossing of the optimal dividing surface. Just as in classical TST, therefore, one needs to locate a dividing surface which eliminates most of the recrossing, and there are some systems (e.g. diffusive reactions) for which no such surface exists, and which therefore cannot be adequately described by quantum TST.

Third, unlike classical TST, quantum TST is not a strict upper bound to the exact rate. This is because, in a quantum system, recrossing of the dividing surface does not necessarily reduce the rate;  real-time coherence effects may in fact increase it, as is happening to a small extent in Fig.~3a. Hence one can only apply quantum TST when it is safe to  assume that real-time coherence has a small effect on the rate, in which case the \shortt limit gives a {\em good approximation} to an upper bound. One can then proceed (almost) as in classical TST, choosing the (approximately)
 optimal dividing surface to be the one that minimises the TST rate, which is what was done in the calculation of Fig.~3a. Moving the dividing surface any appreciable distance away from the (approximately) optimal dividing 
 surface causes the quantum TST rate to become exponentially too large [see Fig.~3c], just as in classical TST.
This restriction that quantum TST is only applicable if real-time coherence effects are small is only to be expected: if coherence effects are large, one has no choice but to compute the real-time quantum dynamics.\cite{fourth}

\subsection{Quantum statistics}
\label{ssec:qmstat}

We now return to why the quantum TST expression in \eqn{eq:kwm} breaks down below temperatures of $2T_{\rm c}$. It is clear that the time-correlation function of \eqn{eq:cfsnf} still gives the exact rate in the $t \to \infty$ limit (see Fig.~3b). The fault in \eqn{eq:cfsnf} is therefore not in the dynamics: it is a failure in the treatment of the quantum statistics.

We can analyse the statistics by discretising imaginary time, such that the statistical part of \eqn{eq:cfsnf} is written as
\begin{align}
\expect{q^\ddag&-\Delta_1/2|e^{-\beta{\hat H}}|q^\ddag+\Delta_1/2}= \nonumber \\
& \int_{-\infty}^\infty\,dq_2\dots \int_{-\infty}^\infty\,dq_N
\expect{q^\ddag-\Delta_{1}/2|e^{-\beta_N{\hat H}}|q_2}\nonumber\\ 
&\times\left[\prod_{i=2}^{N-1}\expect{q_{i}|e^{-\beta_N{\hat H}}|q_{i+1}} \right]\nonumber\\ &\times\expect{q_{N}|e^{-\beta_N{\hat H}}|q^\ddag+\Delta_1/2}
\label{eq:linpol} % Linear Polymer
\end{align}
where $\beta_N=\beta/N$. In the $N \to \infty$ limit, this becomes
\begin{align}
\expect{q^\ddag&-\Delta_1/2|e^{-\beta{\hat H}}|q^\ddag+\Delta_1/2} = \nonumber \\
& \lim_{N\rightarrow\infty}
\left(m\over 2\pi\beta_N\hbar^2\right)^{N/2}\int_{-\infty}^\infty\,dq_2\dots \int_{-\infty}^\infty\,dq_N
\nonumber \\ 
&\times e^{-m(q_{2}-q^\ddag+\Delta_1/2)^2/2\beta_N\hbar^2}e^{-\beta_NV(q^\ddag-\Delta_1/2)/2}\nonumber \\ 
&\times \prod_{i=2}^{N-1}e^{-m(q_{i+1}-q_i)^2/2\beta_N\hbar^2}e^{-\beta_NV(q_i)}\nonumber\\
&\times e^{-m(q_{N}-q^\ddag-\Delta_1/2)^2/2\beta_N\hbar^2}e^{-\beta_NV(q^\ddag+\Delta_1/2)/2}
\end{align}
The right-hand side is a classical partition function for a polymer `string'  connecting the points $q^\ddag \pm \Delta_1/2$.  The quantum statistics is dominated by fluctuations of the polymer string around the minima on its potential energy surface. 

To characterize these minima, we can use what we know about stationary points on the related ring-polymer potential surface. \cite{benders,jor}
In a ring polymer, the stationary points are discretized  periodic orbits on the inverted molecular potential energy surface $-V$,  with period $\beta \hbar$. There is a cross-over temperature $T_{\rm c} = 1/k_{\rm B}\beta_{\rm c} = \hbar\omega_c/2\pi k_{\rm B}$, where $\omega_c$ is the imaginary frequency at the top of the barrier, such that when $T > T_{\rm c}$ the system has insufficient time to complete an orbit and thus the only stationary point is the `trivial orbit', in which the system remains at rest, at the bottom of $-V$ (i.e.\ on the barrier top) for a time $\beta\hbar$; when $T < T_{\rm c}$, the system can complete an orbit, which is referred to as the `instanton'. This cross-over marks the transition between shallow tunnelling, where the system penetrates the parabolic tip of the barrier, and deep tunnelling, where it follows a more delocalised path (and can cut corners\cite{benders,coltrin} etc.). Another way to think of cross-over is that, above $T_{\rm c}$, the springs are too stiff to allow the polymer to stretch over the top of the barrier; below $T_{\rm c}$, the springs are sufficiently weak that the polymer can stretch over the barrier and relax into the geometry corresponding to the discretized instanton orbit. For the special case of a parabolic barrier, there is nothing
in the ring-polymer potential to stop the polymer stretching indefinitely below $T_{\rm c}$, where the rate is therefore undefined. 

For the linear polymer in \eqn{eq:linpol}, the end-points are constrained to be symmetric about $q^\ddag$, so the stationary points are now classical trajectories (on $-V$) connecting the end points in time $\beta \hbar$; they  need not be periodic orbits. Also, the polymer is not a ring, but a string, and so the force constant associated with relaxation into an instanton is half that of a ring-polymer. As a result, the polymer relaxes over the top of the barrier at $T = 2T_{\rm c}$, to form half a periodic orbit (or a `half-instanton'). It is these half-instantons that cause the rate to be negative in the Eckart barrier calculation of Fig.~3b. Figure~4 plots the quantum Boltzmann matrix used to obtain Fig.~3b. The two peaks at $\Delta=\pm 1.7$ a.u.\ are produced by two half-instantons, which are minima on the linear-polymer potential surface.\cite{liudid} It is clear that these half instantons, and the fluctuations around them which dominate the Boltzmann matrix, are spurious. \cite{inst}  For a parabolic barrier, the behaviour of \eqn{eq:kwm} below $T = 2T_{\rm c}$ is even more disastrous, since the rate becomes undefined in the range $2T_{\rm c}>T>T_{\rm c}$ (where it should be finite). \cite{paraprob}

\begin{figure}[tb]
 \resizebox{0.8\columnwidth}{!} {\includegraphics[angle=270]{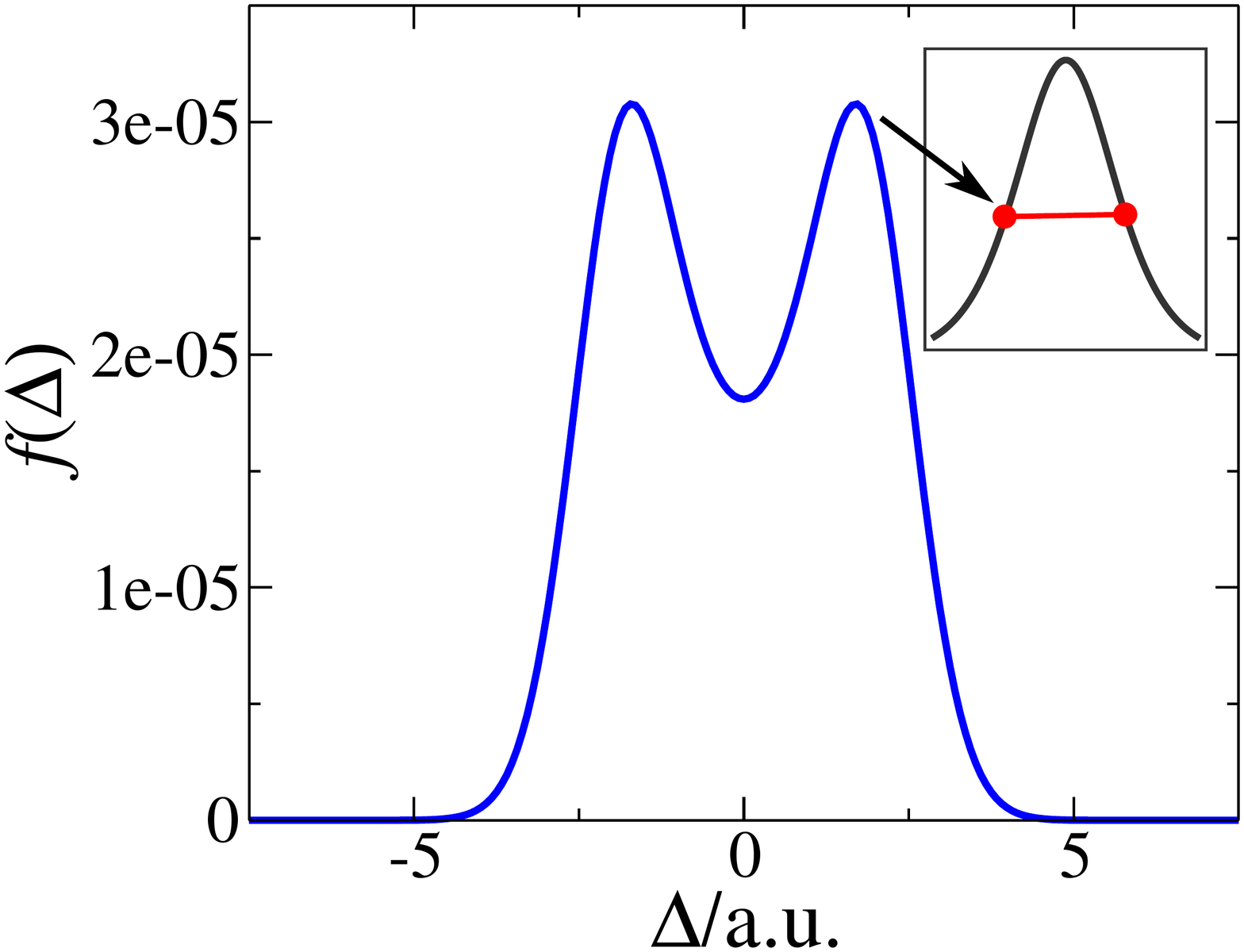}}
 % data from tjhh2@espeto:~/Projects/NewForm/wignerdelta/archive/b3/n1
 \caption{Plot of  $f(\Delta)\!\!={\expect{\!\!\!-\!\Delta/2|e^{-\beta{\hat H}}| \Delta/2}}$, for the Eckart barrier system at $k_{\rm B}\beta=3\times 10^{-3}\ K^{-1}$. Note the bimodal peaks caused by (spurious) half-instantons (red line in inset) extending across the reaction barrier (black curve).}
\end{figure}

These statistical problems are symptoms of a fundamental flaw. In choosing a dividing surface which is a single point we have constrained the quantum Boltzmann distribution in such a way as to make it non-positive-definite. At high temperatures, the errors in the statistics are small. But below $2T_{\rm c}$, the errors are big enough to dominate the TST rate, giving negative or undefined rates as just described. It will become clear below that the reason the statistics is non-positive-definite is that the single-point constraint has destroyed the invariance of the quantum Boltzmann distribution to imaginary-time translation.

\section{Quantum TST with correct statistics}
\label{sec:QTSTstat}
We now construct a general flux-side time-correlation function which is invariant to imaginary-time translation, and which, like \eqn{eq:cfsnf}, uses the same function of path-integral space for the flux and side dividing surfaces. We will find below that this condition gives a quantum TST for which the \shortt quantum statistics is positive-definite, thus avoiding 
 the statistical problems just described. 

\subsection{Generalized Kubo-transformed side-side time-correlation function}
\label{ssec:kubo}

We construct first a side-side time-correlation function which is invariant to imaginary-time translation. We take the quantum partition function for the system, insert the identity of \eqn{eq:iden} at $N$ equally spaced imaginary-time intervals $\beta_N \equiv \beta/N$,  and introduce a dividing surface $f({\bf q})$ which is symmetric under cyclic permutation of the $N$ coordinates ${\bf q}\equiv \{q_i\}$.  This results in the side-side time-correlation function
\begin{align}
C_{\rm ss}^{[N]}(t) = & \int\! d{\bf q}\, \int\! d{\bf z}\,\int\! d{\bf \Delta}\,h[f({\bf q})]h[f({\bf z})]\nonumber\\ 
\times&\prod_{i=1}^{N}\expect{q_{i-1}-\Delta_{i-1}/2|e^{-\beta_N{\hat H}}|q_i+\Delta_i/2}
\nonumber\\ 
\quad\times &\expect{q_i+\Delta_i/2|e^{i{\hat H}t/\hbar}|z_i}
\expect{z_i|e^{-i{\hat H}t/\hbar}|q_i-\Delta_i/2}
\label{eq:cssN} % Css with N beads
\end{align}
where $\int d{\bf q}\equiv \prod_{i=1}^N \int d{q_i}$ etc., and the $i=1\dots N$ are defined cyclically.
%A schematic illustration of $C_{\rm ss}^{[N]}(t)$ is shown in Fig.\ 5 (for the case of N=3).  
We then take the $N\rightarrow\infty$ limit, in which the permutational symmetry of $f({\bf q})$ ensures that $C_{\rm ss}^{[N]}(t)$ is invariant to imaginary-time translation.

Two  examples of functions $f({\bf q})$ that are invariant to cyclic permutation of the coordinates are the centroid dividing surface
\begin{align}
f({\bf q})={\overline q}_0-q^\ddag
\end{align}
where ${\overline q}_0=\sum_{i=1}^N{q_i}/N$ and $q^\ddag$ is a parameter that locates the surface,  and the cone
\begin{align}
f({\bf q})&=\cos\phi\,{\overline q}_0+{\sin\phi \over \sqrt{N}} \,\left|\sum_{j=1}^Ne^{i2\pi j/N}q_j\right|-q^\ddag
\label{eq:cone}
\end{align}
where $\phi$ determines the pitch of the cone. These simple functions will often be sufficient to give a good dividing surface in \eqn{eq:cssN} (see Sec.~\ref{ssec:optdivsurf}), but of course more general permutationally-invariant functions might sometimes need to be used. \cite{braams}

Time-correlation functions with the structure of \eqn{eq:cssN} have not appeared before in the reaction dynamics literature.  However, the interpretation of such a function in the $N \to \infty$ limit is simple: it correlates a property of the entire imaginary-time Feynman path at time $t=0$  with the same property at some later time~$t$. This would seem to be the most general way to construct a quantum time-correlation function.\cite{notun} It is easy to show that the integral would collapse to a standard Kubo-transformed time-correlation function
if the heaviside functions in \eqn{eq:cssN} were replaced by linear operators. We may therefore regard \eqn{eq:cssN} as a generalization of the Kubo transformed time-correlation function to non-linear operators. 

\subsection{Quantum TST}
\label{ssec:QTST}

It is straightforward to obtain the corresponding flux-side time-correlation function and to take the \shortt limit, thus obtaining a \shortt quantum TST in  which the quantum statistics is now invariant to imaginary-time translation. Differentiating $C_{\rm ss}^{[N]}(t)$ with respect to $t$, we obtain\cite{suppl}
\begin{align}
C_{\rm fs}^{[N]}(t)=&\int\! d{\bf q}\, \int\! d{\bf z}\,\int\! d{\bf \Delta}\,{\cal \hat F}[f({\bf q})]h[f({\bf z})]\nonumber\\ 
\times&\prod_{i=1}^{N}\expect{q_{i-1}-\Delta_{i-1}/2|e^{-\beta_N{\hat H}}|q_i+\Delta_i/2}
\nonumber\\ 
\quad\times &\expect{q_i+\Delta_i/2|e^{i{\hat H}t/\hbar}|z_i}
\expect{z_i|e^{-i{\hat H}t/\hbar}|q_i-\Delta_i/2}
\label{eq:cfsnfn} % C flux-side New Form with N beads
\end{align}
with
\begin{align}
{\cal \hat F}[f({\bf q})] &= {1\over 2m}\sum_{i=1}^N\left\{{\hat p_i}{\partial f({\bf q})\over\partial q_i} \hat \delta[f({\bf q})]+ \hat \delta[f({\bf q})]{\partial f({\bf q})\over\partial q_i}{\hat p_i}\right\}
\label{eq:superflux} % The SuperFlux operator
\end{align}
where we employ a similar convention to \eqn{eq:fluxdef} (i.e.\ the first term is inserted between $e^{-\beta_N \hat H}\ket{q_i+\Delta_i/2}$ and $\bra{q_i+\Delta_i/2}e^{i{\hat H}t/\hbar}$ in \eqn{eq:cfsnfn}, with ${\hat p_i}$ acting only on $\ket{q_i+\Delta_i/2}$, and so on). 
\Eqn{eq:cfsnfn} is written out in full in Appendix A. We will refer to ${\cal \hat F}[f({\bf q})]$ as the `ring-polymer flux operator', since it gives the collective flux normal to the ring-polymer dividing surface $f({\bf q})$; this flux is  invariant to imaginary-time translation in the $N \to \infty$ limit. Note that 
\Eqn{eq:cfsnfn} reduces to \eqn{eq:cfsnf} in the special case that 
$N=1$. A schematic illustration of $C_{\rm fs}^{[N]}(t)$ is shown in Fig.~5 (for the case of $N=3$). For a discussion of the $t\rightarrow\infty$ behaviour of $C_{\rm fs}^{[N]}(t)$, see Sec.~IV E.

\begin{figure}[tb]
 \resizebox{0.7\columnwidth}{!} {\includegraphics{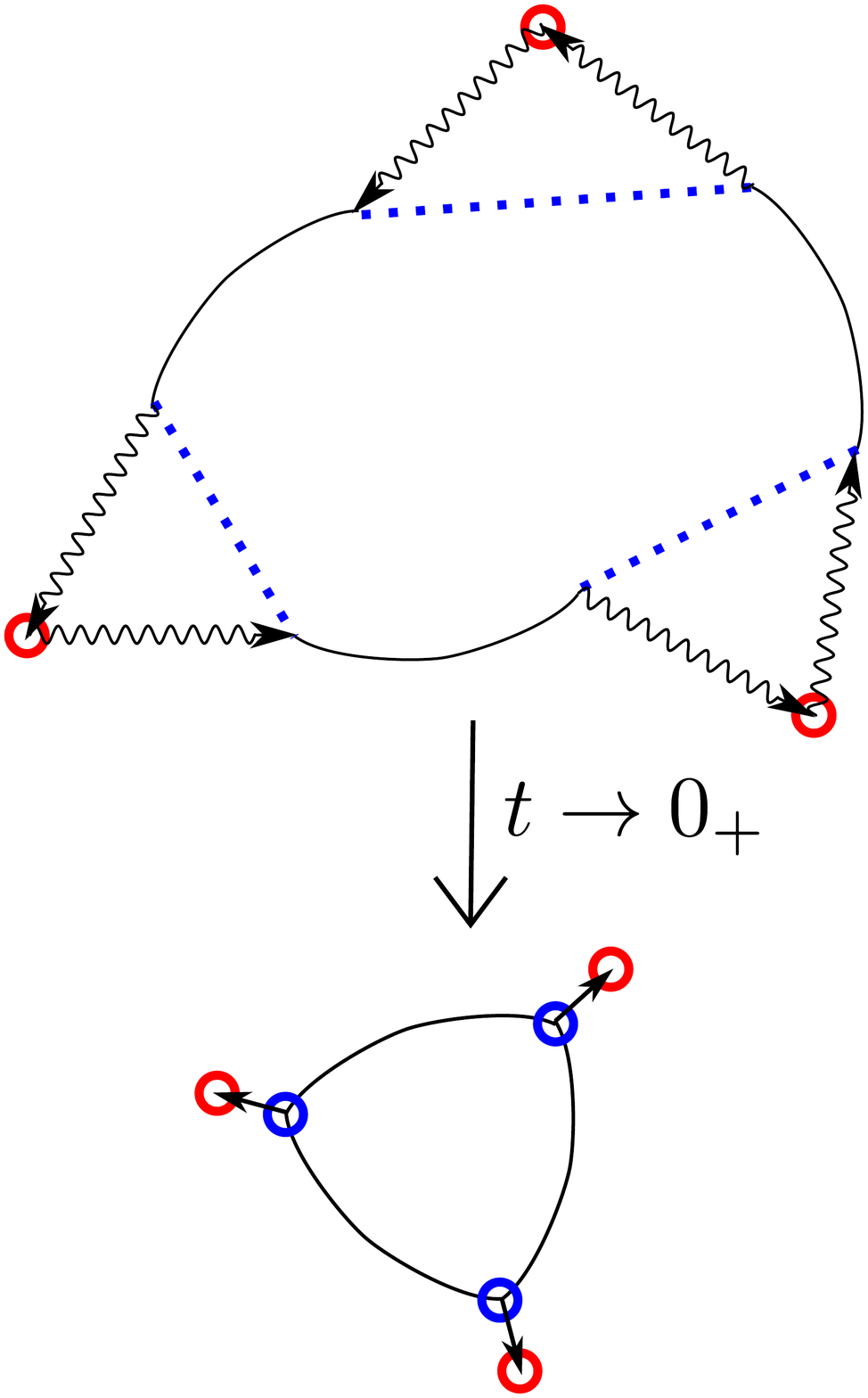}}
 \caption{Schematic representation (following Fig.~2) of the generalized Kubo flux-side time-correlation function
$C_{\rm fs}^{[N]}(t)$ of \eqn{eq:cfsnfn}, for the special case $N=3$. The flux and side dividing surfaces are now permutationally-invariant functions \fq of the initial (blue dashed lines) and final (red circles) ring-polymer coordinates $\bf q$ and $\bf z$. The smaller diagram below shows the effect of taking the \shortt limit.}
\end{figure}

To take the \shortt limit, we use \eqnn{eq:kv}{eq:k} as before, and obtain
\begin{align}\label{biff}
\lim_{t\rightarrow 0_+}&C_{\rm fs}^{[N]}(t)={1\over (2\pi\hbar)^N} \int\! d{\bf q}\, \int\! d{\bf p}\, \int\! d{\bf \Delta}\,\nonumber \\
& \times  
\delta[f({\bf q})] \,{S({\bf q},{\bf p})}h\!\left[f({\bf q}_t)\right]
\nonumber\\ &
\times\prod_{i=1}^{N}\expect{q_{i-1}-\Delta_{i-1}/2|e^{-\beta_N{\hat H}}|q_i+\Delta_i/2}e^{ip_i\Delta_i/\hbar}
\end{align}
where
\begin{align}
S({\bf q},{\bf p})={1\over m}\sum_{i=1}^N{\partial f({\bf q})\over \partial q_i}p_i
\label{eq:sdef}
\end{align}
and ${\bf q}_t={\bf q} + {\bf p}t/m$. This limit is time-independent and non-zero because the flux and side dividing-surfaces are the same; i.e.
\begin{align}
\lim_{t\rightarrow 0_+}f({\bf q}_t)=f({\bf q}) + S({\bf q},{\bf p})t
\end{align}
and hence
  \begin{align}\label{oof}
 \lim_{t\rightarrow 0_+}\delta[f({\bf q})] h\!\left[f({\bf q}_t)\right]=\delta[f({\bf q})] h\!\left[S({\bf q},{\bf p})\right].
   \end{align}
We can therefore define a quantum TST as the limit
  \begin{align}
  k_{Q}^\ddag(\beta)Q_{\rm r}(\beta)=\lim_{t\rightarrow 0_+}\lim_{N\rightarrow\infty}C_{\rm fs}^{[N]}(t)
 \end{align} 
Substituting \eqn{oof} into \eqn{biff}, we obtain
\begin{align}
 k_{Q}^\ddag(\beta)&Q_{\rm r}(\beta)=\lim_{N\rightarrow\infty}{1\over (2\pi\hbar)^N}\int\! d{\bf q}\, \int\! d{\bf p}\,\int\! d{\bf \Delta}\, \nonumber \\
& \times \delta[f({\bf q})] \,{S({\bf q},{\bf p})}h\!\left[S({\bf q},{\bf p})\right]
\nonumber\\ 
& \times \prod_{i=1}^{N}\expect{q_{i-1}-\Delta_{i-1}/2|e^{-\beta_N{\hat H}}|q_i+\Delta_i/2}e^{ip_i\Delta_i/\hbar}
\label{eq:knfn} % k of the New Form with N beads
\end{align}
This expression is an $N$-bead generalization (in the limit $N\rightarrow\infty$) of \eqn{eq:kwm}, and can be regarded as an $N$-fold generalization of a Wigner transform (which was obtained by
 taking the \shortt limit and thus linearizing exactly the dynamics at
 $N$ imaginary-time intervals in the Boltzmann operator). \cite{shortcut}

\subsection{Positive-definite quantum statistics}
\label{ssec:pdqmstat}

We now prove  that the invariance of the dividing surface \fq to imaginary time-translation (in the \largen limit) ensures that \eqn{eq:knfn} gives positive-definite quantum statistics at all temperatures. This proof is not too difficult because, as we now show, all  $N$ of the Fourier transforms can be evaluated analytically. (This is another advantage over \eqn{eq:kwm}, in which the single Fourier transform must usually be evaluated numerically.) 
 The trick is to apply the ${\bf q}$-dependent orthogonal coordinate transformation
\begin{align}
({\bf p},{\bf \Delta})\rightarrow ({\bf P},{\bf D})
\label{eq:ct1} % Co-ordinate Transformation 1
\end{align} 
to the integrand in  \eqn{eq:knfn}, where
 \begin{align}
P_j({\bf q}) =& \sum_{i=1}^N p_iT_{ij}({\bf q}) \nonumber\\D_j({\bf q}) =& \sum_{i=1}^N \Delta_iT_{ij}({\bf q}) 
,\quad\quad\quad j=0\rightarrow N-1
\label{eq:ct2} % Co-ordinate Transformation 2
\end{align}
 and 
\begin{align}\label{bastard}
T_{i0}({\bf q}) = {1\over \sqrt{B_N({\bf q})}}{\partial f({\bf q})\over \partial q_i}
\end{align}
with
\begin{align}
B_N({\bf q})=\sum_{i=1}^N\left[{\partial f({\bf q})\over \partial q_i}\right]^2
\end{align}
and the other elements of ${\bf T(q)}$ can be chosen in any way that makes ${\bf T(q)}$ orthogonal.
The variable $P_0({\bf q})$ describes the momentum perpendicular to the dividing surface \fq at the point ${\bf q}$.
On applying this transformation, we obtain an expression containing $N\!-\!1$ Dirac-delta functions in $D_j$, 
($j=1\rightarrow N\!-\!1$), which can be integrated out to give
\begin{align}
&k_{Q}^\ddag(\beta)Q_{\rm r}(\beta)={1\over (2\pi\hbar)^N}\int\! d{\bf q}\, \int\! dP_0\,\int\! dD_0\,\nonumber \\
& \times \delta[f({\bf q})] \,\sqrt{B_N({\bf q})} \frac{P_0}{m}\ h\!\left(P_0\right)e^{iP_0D_0/\hbar}
\nonumber\\ 
& \times\prod_{i=1}^{N}\expect{q_{i-1}-T_{i\!-\!1\,0}({\bf q})D_0/2|e^{-\beta_N{\hat H}}|q_i+T_{i0}({\bf q})D_0/2}
\label{eq:knm} % k after Normal Mode transformation
\end{align}
This equation resembles a symmetrised version of \eqn{eq:kwm}, with only one Fourier transform involving a `stretching' variable $D_0$. However, in contrast to \eqn{eq:kwm}, where the stretch $\Delta$ is concentrated at one point in the ring-polymer loop, $D_0$ describes a concerted stretch which is smeared out gradually around the loop, such that the separation between adjacent ring-polymer sections tends to zero in the $N\to\infty$ limit (since normalisation ensures that $T_{i0} \sim N^{-1/2}$). 

In Appendix B, we prove that \eqn{eq:knm} simplifies further. First, the smearing-out of $D_0$ around the loop ensures that the potential energy terms $V[q_i\pm T_{i0}({\bf q})D_0/2]$ become independent of $D_0$ in the $N\to \infty$ limit. Second, the  invariance of $f({\bf q})$ to imaginary-time translation ensures that there are no cross-terms between $D_0$ and ${\bf q}$. As a result, \eqn{eq:knm} simplifies to
\begin{align}
k_{\rm Q}^\ddag(\beta)&Q_{\rm r}(\beta)=\lim_{N\rightarrow\infty}{1\over (2\pi\hbar)^N}\int\! d{\bf q}\, \int\! dP_0\,\int\! dD_0\, \nonumber \\
& \times \delta[f({\bf q})] \,\sqrt{B_N({\bf q})}\frac{P_0}{m}h\!\left(P_0\right) \nonumber\\ 
& \times e^{-mD_0^2/2\beta_N\hbar^2}e^{iP_0D_0/\hbar}\prod_{i=1}^{N}\expect{q_{i-1}|e^{-\beta_N{\hat H}}|q_i}
\label{eq:kgd} % k with a Gaussian in D
\end{align}
i.e.\ the Boltzmann distribution has a simple Gaussian dependence on just the one variable $D_0$. 
It is therefore positive-definite at all temperatures, and incapable of giving the sort of bimodal dependence
(seen in Fig.~4) that causes \eqn{eq:kwm} to break down at $T<2T_{\rm c}$.

\subsection{Emergence of RPMD-TST}
\label{ssec:emergence}

We can simplify this expression still further by integrating over $D_0$, which has the effect of joining up the $N$ sections of the ring-polymer into a continuous loop:
\begin{align}
k_{Q}^\ddag(\beta)&Q_{\rm r}(\beta) =\lim_{N\rightarrow\infty}{1\over (2\pi\hbar)^N}\int\! d{\bf q}\, \int\! dP_0\, \nonumber \\
& \times \delta[f({\bf q})] \,\sqrt{B_N({\bf q})}\frac{P_0}{m}h\!\left(P_0\right)
\nonumber\\ 
& \times\sqrt{2\pi\beta_N\hbar^2\over m}e^{-P_0^2\beta_N/2m}\prod_{i=1}^{N}\expect{q_{i-1}|e^{-\beta_N{\hat H}}|q_i}
\end{align}
We then make the substitution
 \begin{align}
\lim_{N\rightarrow\infty}&\expect{q_{i-1}|e^{-\beta_N{\hat H}}|q_i} = \sqrt{m\over 2\pi\beta_N\hbar^2} \nonumber \\
& \times e^{-m(q_i-q_{i-1})^2/2\beta_N\hbar^2}e^{-\beta_N [V(q_i)+V(q_{i-1})]/2}
 \end{align}
and convert $N\!-\!1$ of the normalisation factors $\sqrt{m/2\pi\beta_N\hbar^2}$ into integrals over $\exp(-P_j^2\beta_N/2m)$, where $j=1\rightarrow N-1$. Transforming back from $\bf P$ to $\bf p$, we obtain
\begin{align}
k_{Q}^\ddag(\beta)& Q_{\rm r}(\beta)= \lim_{N\rightarrow\infty}{1\over (2\pi\hbar)^N}\int\! d{\bf q}\, \int\! d{\bf p} \nonumber \\
& \times \delta[f({\bf q})] \, {S({\bf q},{\bf p})}h\!\left[S({\bf q},{\bf p})\right] \nonumber\\ 
& \times \prod_{i=1}^{N}e^{-m(q_i-q_{i-1})^2/2\beta_N\hbar^2}e^{-\beta_N V(q_i)}e^{-p_i^2\beta_N/2m}
\label{eq:krpmdtst}
\end{align}
where $S({\bf q},{\bf p})$ is defined as in \eqn{eq:sdef}. 

\Eqn{eq:krpmdtst} is identical to the $N \to \infty$ limit of a classical TST expression, in which the reciprocal temperature is $\beta_N$, the dividing surface is $f({\bf q})$,  and the Hamiltonian is
\begin{align}
H({\bf q},{\bf p}) = \sum_{i=1}^N {p_i^2 \over 2 m}+
\sum_{i=1}^N\left[{m(q_i-q_{i-1})^2\over 2(\beta_N\hbar)^2}+V(q_i)\right]
\end{align}
Remarkably, these are precisely the conditions that define RPMD-TST.  We can thus write
\begin{align}
k_{\rm Q}^\ddag(\beta) \equiv  k^{\ddag}_{\rm RPMD}(\beta)
\end{align} 
where it is understood that the RPMD calculation uses $N\rightarrow\infty$ beads (which in practice means that $N$ is increased until numerical convergence is reached) and that the RPMD dividing surface $f({\bf q})$ is invariant under cyclic permutation of the polymer beads. 

Hence in deriving \eqn{eq:krpmdtst},  we have found that RPMD rate-theory, which was originally
 conceived using heuristic arguments\cite{rates,refined} (and for which the best justification until now was that it interpolated correctly between various limits\cite{jor}) is  the unique \cite{notun} \shortt quantum TST that gives positive-definite statistics.
    
\subsection{Long-time behaviour and choice of optimal dividing-surface}
\label{ssec:optdivsurf}

In general, $C_{\rm fs}^{[N]}(t)$ of \eqn{eq:cfsnfn} does not give the exact quantum rate in the \longt limit, but must be added to correction terms which can only be derived in closed form in special cases. However, we will prove in Part II that these correction terms are {\em zero} if there is no recrossing of $f({\bf q})$ (or of surfaces perpendicular to $f({\bf q})$ in ring-polymer space) and hence that 
$k^\ddag_{\rm Q}(\beta)$ gives the exact quantum rate  $k_{\rm Q}(\beta)$ in the absence of recrossing.  In this respect, the relation between $k^\ddag_{\rm Q}(\beta)$ and $k_{\rm Q}(\beta)$ is thus the same as that between $k^\ddag_{\rm C}(\beta)$ and $k_{\rm C}(\beta)$ in classical TST. However, unlike classical TST, $k^\ddag_{\rm Q}(\beta)$ does not give a strict upper bound to $k_{\rm Q}(\beta)$. That this is so is evident from previous benchmark comparisons of RPMD-TST, where $k^\ddag_{\rm Q}(\beta)$ was found\cite{jor}  to underestimate the rates for symmetric barriers below cross over. 

Hence, just as with $k^\ddag_{\rm WM}(\beta)$ of Sec.~III, one can apply $k^\ddag_{\rm Q}(\beta)$ only if real-time quantum coherence effects
are small, in which case $k^\ddag_{\rm Q}(\beta)$ gives a {\em good approximation} to an upper bound to $k_{\rm Q}(\beta)$. One can then proceed 
exactly as in a practical RPMD calculation, by taking the (near) optimal dividing surface to be the surface that maximises the free-energy of the ring-polymer ensemble.  The restriction that \fq be invariant under cyclic permutation of the beads is not an additional constraint, since a little thought shows that any dividing surface that maximises the ring-polymer free energy must necessarily have this property.

From the extensive previous work on RPMD \cite{jor,rates,refined,azzouz,bimolec,ch4,mustuff,anrev,yury,tommy1,tommy2,
 stecher} and centroid-TST, \cite{gillan1,gillan2,centroid1,centroid2,schwieters} therefore, we already know a lot about the likely form of the optimal dividing surface in $k_{\rm Q}^\ddag(\beta)$. It is worth summarising these findings here. For a parabolic barrier or free particle, the optimal dividing surface is the centroid; in fact we will show (in Part II) that the centroid-dividing surface
makes $C_{\rm fs}^{[N]}(T)$  time-independent and equal to the exact rate for these systems. For real systems, there are two tunnelling regimes separated by a cross-over temperature $T_{\rm c}$ (see Sec.~\ref{ssec:qmstat}): for $T>T_{\rm c}$, the statistics is dominated by fluctuations around a point, and so the centroid is usually also a good dividing surface; for $T<T_{\rm c}$, the centroid is also a good dividing surface, provided the temperature is not too low, and provided the barrier is reasonably symmetric; for a strongly asymmetric barrier, however, centroid-TST breaks down,\cite{jor,schwieters} since it is necessary to let the ring polymers stretch over the barrier, using a cone-like dividing surface such as \eqn{eq:cone}. Hence centroid-TST often works, but is a special case of RPMD-TST, which Secs.~\ref{ssec:kubo} -- \ref{ssec:emergence} have shown arises naturally as the most general form of \shortt flux which gives positive-definite quantum statistics.
 
\subsection{Numerical considerations}
\subsubsection{Value of $N$ needed to give good quantum statistics}
\label{ssec:convqm}

In Sec.~\ref{ssec:emergence} we showed that the expression given in \eqn{eq:knfn} for $k_{\rm Q}^\ddag(\beta)$ simplifies to the RPMD-TST expression of \eqn{eq:krpmdtst}, and that this occurs because the dividing surface \fq becomes invariant to imaginary-time-translation in the \largen limit.  However, in a practical calculation $N$ will be finite, and treated as a numerical convergence parameter. The question then arises as to how big $N$ needs to be to ensure that \eqn{eq:knfn} gives sufficiently good numerical agreement with \eqn{eq:krpmdtst}, indicating that this value of $N$ gives an \fq which, although not perfectly invariant to imaginary-time translation, is sufficiently close to it. In a practical RPMD calculation, $N$ is often quite large (e.g. $N\!=\!10$--100 is needed to converge the rate for a hydrogen-transfer-type reaction at temperatures not too far below cross-over;\cite{jor} $N\!=\!1000$ for electron-transfer reactions\cite{tommy2}). However, given that just $N$=1 is sufficient to achieve a good approximation to positive-definite quantum statistics for $T>2T_{\rm c}$, it is likely that most of the beads are needed to evaluate numerically the Boltzmann matrix, and that the value of $N$ needed to give a good approximation to positive-definite quantum statistics is much smaller.

This turns out to be the case. Table I shows a comparison between numerical evaluations of equation \eqn{eq:knfn}  (done using quantum wave-packet methods) and the RPMD-TST equation \eqn{eq:krpmdtst} (done using Monte Carlo sampling of ring-polymer space). The value of $N$ needed in \eqn{eq:knfn} to obtain good numerical agreement with \eqn{eq:krpmdtst} is still only $N\!=\!4$ at a temperature of $k_{\rm B}\beta \!=\!4 \times 10^{-3}$K$^{-1}$, which is significantly below the cross-over temperature of $k_{\rm B}\beta_{\rm c}\!=\!2.69\times 10^{-3}$K$^{-1}$. Hence, as expected, most of the beads used to evaluate the RPMD-TST expression \eqn{eq:krpmdtst} are required to evaluate the Boltzmann operator numerically, with only a few being required to give a good approximation to positive-definite quantum statistics. We should emphasise that, despite needing many more beads, the ring-polymer expression \eqn{eq:krpmdtst} is of course vastly more efficient to evaluate than \eqn{eq:knfn}.

\begin{table*}
\begin{ruledtabular}
\begin{tabular}{l|l|l|l|l|l|l}
%\begin{tabular}[h]{l|l|l|l|l|l|l}
$k_{\rm B}\beta/10^{-3}$K$^{-1}$ & $N=1$ & $N=2$ & $N=3$ & $N=4$ & RPMD ($N$) & QM\\ \hline
1       & 13.26 & 13.49 & 13.50 & 13.50 & 13.51 (8)  & 14.15 \\
2       & 8.76\tp{-2} & 0.1336 & 0.1342 & 0.1343 & 0.136 (8)  & 0.1501  \\ 
3       & -1.026\tp{-2} & 2.794\tp{-3} & 2.804\tp{-3} & 2.811\tp{-3}& 2.81\tp{-3} (32)  & 3.485\tp{-3} \\
4       & -5.701\tp{-3} & 1.467\tp{-4} & 1.409\tp{-4} &  1.430\tp{-4} & 1.42\tp{-4} (256) & 2.044\tp{-4}
\end{tabular}
\end{ruledtabular}
\caption{Comparison of \shortt quantum TST rates computed for the Eckart barrier system (see text)
 using \eqn{eq:knm}, with $N$=1--4,  and using the RPMD-TST expression of \eqn{eq:krpmdtst}, with a value of $N$ sufficient to converge the results to within 0.5\%. The exact quantum results (QM) are also shown.}
\end{table*}

\subsubsection{Comparison with the LSC-IVR method}
There is evidently a close link between the RPMD-TST and  LSC-IVR\cite{scivr1,scivr2,scivr3} methods, since both are consistent with linearization approximations to an exact flux-side time-correlation function. In RPMD-TST, one linearizes $C_{\rm fs}^{[N]}(t)$
and takes the \shortt limit (in which the lineariztion is exact), and in the LSC-IVR method one linearizes
 $C_{\rm fs}(t)$, and propagates the linearized trajectories for a finite time $t>0$. Clearly both methods  ignore the effects of real-time quantum coherence, and thus we can expect them to make predictions of the rate that are of comparable accuracy.
A preliminary comparison of the results in Table~I with those of ref.\ \onlinecite{scivr3} suggest that this is indeed the case. For example, both methods make predictions that are within 20-30\% of the exact quantum rate at temperatures of $T\!=\!250$K and 333K. More systematic comparisons of the two methods would be a good subject for future research.

\section{Generalization to multi-dimensions}
\label{sec:multid}

The treatment of Sec.~\ref{sec:QTSTstat} generalizes immediately to multi-dimensional systems, and we now summarise how this can be done.  

Without loss of generality, we can represent the space of an $F$-dimensional system using cartesian coordinates $q_j$, $j=1\rightarrow F$, with a mass $m_j$ associated with coordinate $q_j$. We can also define ring-polymer coordinates 
${\bf q}\equiv \{q_{i,j}\}$, and analogous generalizations of $\bf p$, $\bf \Delta$, etc.,
where each $i=1\rightarrow N$ labels a different replica of the system.

We can then construct the multi-dimensional version of $C_{\rm fs}^{[N]}(t)$ by inserting $N$ identities analogous to \eqn{eq:iden} into the quantum partition function, introducing a dividing surface \fq, which is invariant under {\em collective} cyclic permutations of the coordinates ${\bf q}$ (i.e.\  cyclic permutation among the $N$  replicas), and differentiating with respect to time. This gives an expression analogous to \eqn{eq:cfsnfn}, with the replacements 
\begin{align}
\ket{q_i+\Delta_i/2}\rightarrow \ket{q_{i,1}+\Delta_{i,1}/2,\dots ,q_{i,F}+\Delta_{i,F}/2 }
\end{align}
and so on, and in which the ring-polymer flux operator  is 
\begin{align}
& {\cal \hat F}[f({\bf q})]=\nonumber \\
& \times \sum_{j=1}^F{1\over 2m_j}\sum_{i=1}^N\left\{{\hat p_{i,j}}{\partial f({\bf q})\over\partial q_{i,j}} \hat \delta[f({\bf q})]+ \hat \delta[f({\bf q})]{\partial f({\bf q})\over\partial q_{i,j}}{\hat p_{i,j}}\right\}
\end{align}
and is inserted around the ring in a manner analogous to ${\cal \hat F}[f({\bf q})]$ of \eqn{eq:superflux}.

The rest of the derivation in Sec.~\ref{sec:QTSTstat} can then be followed step-by-step. We take the \shortt limit of $C_{\rm fs}^{[N]}(t)$,
 to obtain an expression analogous to \eqn{biff} in which $S({\bf q},{\bf p})$ is defined according to
\begin{align}
S({\bf q},{\bf p})=\sum_{j=1}^F{1\over m_j}\sum_{i=1}^N{\partial f({\bf q})\over \partial q_{i,j}}p_{i,j}
\end{align}
We then carry out a coordinate transformation analogous to \eqnn{eq:ct1}{eq:ct2}, in which ${P}_0({\bf q})$ is  the momentum orthogonal to the dividing surface $f({\bf q})$, and follow the rest of the derivation in Sec.~\ref{sec:QTSTstat} and Appendix B. As in Sec.~\ref{ssec:pdqmstat}, the imaginary-time-translation invariance of \fq in the \largen limit is essential in arriving at the final result, which is simply RPMD-TST in multiple dimensions.  

%Note also that the proof in Part II that the \longt behaviour of $C_{\rm fs}^{[N]}(t)$ gives the exact quantum
%rate if there is no recrossing of \fq (or of surfaces perpendicular to it in ring-polymer s%pace) applies also to multi-dimensional systems. 

\section{Conclusions}
\label{sec:con}

The results of this article are surprising, since it has been widely assumed that a true quantum TST,  in the form of a non-zero \shortt limit of a quantum flux-side time-correlation function,  does not exist. This article
has shown that such a limit does exist (and the forthcoming Part II will show that it gives the exact quantum rate in the absence of recrossing). Remarkably, the \shortt limit gives expressions for the rate which are identical to two results obtained previously as educated guesses. One of these is Wigner's expression,\cite{mill,wiggie} which we found is indeed a \shortt quantum TST, but which gives bad (i.e.\ non-positive-definite) quantum statistics at low temperatures. The other is RPMD-TST, in which the use of a dividing surface which is invariant to imaginary-time translation guarantees that the \shortt limit gives positive-definite quantum statistics. As a result, RPMD-TST appears to be the {\em unique}\cite{notun}  \shortt quantum TST that gives the correct quantum statistics at all temperatures.

Unlike classical TST, RPMD-TST does not give a strict upper bound to the exact quantum rate. Instead, it gives a {\em good approximation} to an upper bound, {\em provided} real-time quantum coherence does not significantly affect the rate.  In such cases, the optimal dividing surface is {\em close to} the surface that maximises the free-energy of the ring-polymers. In the shallow-tunnelling regime, this surface is the centroid, which explains why centroid-TST\cite{gillan1,gillan2,centroid1,centroid2} (i.e.\ RPMD-TST using a centroid dividing surface) works in this regime. Centroid-TST also works below the cross-over to deep tunnelling if the barrier is symmetric, but breaks down for asymmetric barriers, since the optimal dividing surface then involves ring-polymer stretching coordinates. In the high-temperature limit, 
 quantum coherence vanishes, and  classical TST emerges from RPMD-TST as a special limiting case.

To derive RPMD-TST,  we used the property that the linearization approximation \cite{scivr1,scivr2,scivr3} of the flux-side time-correlation function is exact in the $t\rightarrow 0_+$ limit, where the dynamics  can be represented (without approximation) by an ensemble of classical free particles,  with initial positions distributed along  (imaginary-time) Feynman paths. We showed that the standard flux-side time-correlation function correlates the initial flux of a particle located at a single point on any given path, with the side of a particle located at a {\em different} point (on the same path). In other words, the flux and side dividing surfaces are {\em different} functions of path-integral space, which is why the time-correlation function tends smoothly to zero as $t\rightarrow 0_+$. We then introduced a generalized Kubo-transformed flux-side time-correlation function, which allows one to use the {\em same} flux and side dividing surfaces; it is this property that gives the non-zero \shortt  limit (i.e.\ RPMD-TST).

In a practical RPMD simulation,\cite{rates,refined,azzouz,bimolec,ch4,mustuff,anrev,yury,tommy1,tommy2,stecher} one  does not usually compute the RPMD-TST rate directly;  instead, one computes the
exact classical rate in the extended (fictitious) space of the ring-polymers, thus obtaining a lower bound to the RPMD-TST rate. This exact RPMD rate will be a good approximation to the exact quantum rate provided the TST approximation
 holds good in both the (true, physical) quantum space and the (fictitious)
 classical ring-polymer space. Hence, provided real-time quantum coherence is not important, and provided the TST assumption (that the reaction is dominated by a free-energy bottleneck) holds, we can be confident that an RPMD simulation will give a good approximation to the exact quantum rate.

\begin{acknowledgments}
TJHH is supported by a Project Studentship from the UK Engineering and
Physical Sciences Research Council. 
\end{acknowledgments}

\section*{APPENDIX A} % With x and y instead of q and delta
\label{app:a}
\renewcommand{\theequation}{A\arabic{equation}}
\setcounter{equation}{0}

\Eqn{eq:cfsnfn} written out in full is
\begin{align}\label{mum}
& C_{\rm fs}^{[N]}(t) ={1\over 2m}\int\! d{\bf q}\, \int\! d{\bf z}\,\int\! d{\bf \Delta}\,h[f({\bf z})]\nonumber\\ 
& \times \sum_{i=1}^{N} \Bigg[ \expect{q_{i-1}-\Delta_{i-1}/2|e^{-\beta_N{\hat H}}|q_i+\Delta_i/2} \nonumber\\ 
& \times  \Big\{ {\hat L}_i({\bf q})\expect{q_i+\Delta_i/2| e^{i{\hat H}t/\hbar}|z_i} \expect{z_i|e^{-i{\hat H}t/\hbar}|q_i-\Delta_i/2} \nonumber \\
& + \expect{q_i+\Delta_i/2|e^{i{\hat H}t/\hbar}|z_i} \expect{z_i|e^{-i{\hat H}t/\hbar}|q_i-\Delta_i/2}{\hat R}_i({\bf q}) \Big\} \nonumber \\
& \times \prod_{j=1,j\neq i}^{N}\expect{q_{j-1}-\Delta_{j-1}/2|e^{-\beta_N{\hat H}}|q_j+\Delta_j/2}
\nonumber\\ 
& \times \expect{q_j+\Delta_j/2|e^{i{\hat H}t/\hbar}|z_j}
\expect{z_j|e^{-i{\hat H}t/\hbar}|q_j-\Delta_j/2} \Bigg]
\end{align}
with
 \begin{align}
 {\hat L}_i({\bf q})  =   \hat p_i\, \hat \delta_i[f({\bf q})] \frac{\partial f({\bf q})}{\partial q_i}\nonumber\\ {\hat R}_i({\bf q}) = \frac{\partial f({\bf q})}{\partial q_i}\hat \delta_i[f({\bf q})]\,\hat p_i
  \end{align}
  and where we employ the convention that the $\hat p_i$ operator in  ${\hat L}_i({\bf q})$ acts only on the neighbouring ket $\ket{q_i+\Delta_i/2}$, and the $\hat p_i$ operator in ${\hat R}_i({\bf q})$  acts only on the neighbouring bra $\bra{q_i-\Delta_i/2}$.
Note that there are a variety of other ways to include the flux operators (e.g.\ 
replacing each ${\hat L}_i({\bf q})$ and ${\hat R}_i({\bf q})$ in \eqn{mum} by $[{\hat L}_i({\bf q})+{\hat R}_i({\bf q})]/2$),  all of which are 
equivalent to \eqn{mum}. 
\section*{APPENDIX B}
\label{app:b}
\renewcommand{\theequation}{B\arabic{equation}}
\setcounter{equation}{0}

To prove the link between \eqnn{eq:knm}{eq:kgd}, we first write
\begin{align}
\lim_{N\rightarrow\infty}&\prod_{i=1}^{N}\expect{q_{i-1}-T_{i\!-\!1\,0}({\bf q})D_0/2|e^{-\beta_N{\hat H}}|q_i+T_{i0}({\bf q})D_0/2}\nonumber\\ 
&=\lim_{N\rightarrow\infty}\left({m\over2\pi \beta_N\hbar^2}\right)^{N/2}\nonumber \\
& \times \prod_{i=1}^{N}
e^{-m\left\{q_i-q_{i-1}+D_0[T_{i\!-\!1\,0}({\bf q})+T_{i0}({\bf q})]/2\right\}^2/2\beta_N\hbar^2}\nonumber\\
&\times e^{-\beta_N\left\{V[q_{i}-T_{i0}({\bf q})D_0/2]+V[q_{i}+T_{i0}({\bf q})D_0/2] \right\}/2}\label{eq:a6}
\end{align}
We then note that
\begin{align}
T_{i0}\sim N^{-\frac{1}{2}}
\end{align}
and hence that
\begin{align}
{\beta_N\over 2} \left\{V[q_{i}+T_{i0}({\bf q})D_0/2]+V[q_{i}-T_{i0}({\bf q})D_0/2]\right\} \nonumber\\= \beta_N [V(q_i)+\mathcal{O}(D_0^2 N^{-1})]
\end{align}
which gives 
\begin{align}
\lim_{N\rightarrow\infty} & e^{-\beta_N\left\{V[q_{i}-T_{i0}({\bf q})D_0/2]+V[q_{i}+T_{i0}({\bf q})D_0/2] \right\}/2}\nonumber \\ 
& = e^{-\beta_N\left[V(q_{i-1})+V(q_{i}) \right]/2}
\label{eq:potterm} % Potential term
\end{align}
In other words, the $D_0$-dependence disappears from the potential terms in the $N\rightarrow\infty$ limit, where 
$D_0$ describes a `breathing' mode involving the  collective opening and closing of a sequence of infinitesimal gaps distributed continuously around the ring-polymer loop.

To deal with the kinetic terms, we assume that the dividing surface $f({\bf q})$ has been chosen to be a smooth function of path integral space, such that 
\begin{align}
T_{i0}({\bf q})=T(\tau_i,{\bf q})  
\end{align}
where $T(\tau,{\bf q})$ is a smooth function of the imaginary time $\tau$, evaluated at  $\tau_i = (i-1)\beta_N\hbar$. We then expand $T_{i\pm1\,0}({\bf q})$ as
\begin{align}
T_{i\pm1\,0}({\bf q})=T_{i0}({\bf q}) \pm \beta_N\hbar  {\dot T}_{i0}({\bf q})+\mathcal{O}(\beta_N^2)
\label{eq:texpand} % T expanded
\end{align}
where ${\dot T}_{i0}({\bf q}) =\left.{\partial T(\tau,{\bf q})/ \partial \tau}\right|_{\tau=\tau_i}$, and expand the square in the third line of \eqn{eq:a6} into three terms. The first term gives the usual ring-polymer expression
\begin{align}
\prod_{i=1}^N e^{-m(q_i-q_{i-1})^2/2\beta_N\hbar^2};
\label{eq:rpt} % Ring Polymer Term
\end{align}
the second term gives
\begin{align}
\lim_{N\rightarrow\infty}\prod_{i=1}^N e^{-mD_0^2[T_{i\!-\!1\,0}({\bf q})+T_{i0}({\bf q})]^2/8\beta_N\hbar^2}=e^{-mD_0^2/2\beta_N\hbar^2}
\end{align}
where the simplification on the righthand side is easily shown to result from \eqn{eq:texpand};  the third term gives
\begin{align}
\prod_{i=1}^N e^{-m(q_i-q_{i-1})D_0[T_{i\!-\!1\,0}({\bf q})+T_{i0}({\bf q})]/2\beta_N\hbar^2 }=e^{-g({\bf r})D_0/\hbar}\label{bugger}
\end{align}
where
\begin{align}
g({\bf r}) &= {m\over2\beta_N\hbar}\sum_{i=1}^N (q_{i+1}-q_{i-1})T_{i0}({\bf q})%\nonumber\\
%&= {1\over2\beta_N\hbar}\sum_{i=1}^N (q_{i+1}-q_{i-1}){\partial f({\bf q})\over \partial q_i}
\end{align}
Now, the dividing surface is invariant under imaginary-time translation, i.e.
\begin{align}\label{perms}
{\mathcal P}_{i\rightarrow i+1}f({\bf q})=f({\bf q})
\end{align}
where ${\mathcal P}_{i\rightarrow i+1}$ denotes a cyclic permutation of the ring-polymer beads, such that
 $q_i\rightarrow q_{i+1}$. Since ${q_i-q_{i-1}}\rightarrow 0$ as $N\rightarrow\infty$, it follows that
\begin{align}
\lim_{N\rightarrow\infty} & ({\mathcal P}_{i\rightarrow i+1}-{\mathcal P}_{i\rightarrow i-1})f({\bf q})\nonumber \\
& = \lim_{N\rightarrow\infty} \sum_{i=1}^N(q_{i+1}-q_{i-1}){\partial f({\bf q})\over \partial q_i}\nonumber \\
& = 0
\end{align}
Using \eqn{bastard} to replace $\partial f({\bf q})/ \partial q_i$ by $T_{i0}({\bf q})$, we obtain
\begin{align}
\lim_{N\rightarrow\infty} g({\bf r}) = 0.
\end{align}
Hence the term given in \eqn{bugger}
 tends to unity as $N\rightarrow\infty$, with the result that \eqn{eq:knm} gives \eqn{eq:kgd}. 
Note that this would not have happened if \eqn{perms} did not hold. The imaginary-time-translational invariance of $f({\bf q})$ is therefore {\em essential} for \eqn{eq:knm} to reduce to \eqn{eq:kgd}.


\begin{thebibliography}{10}
\bibitem{beloved}  By which we mean the version of TST discussed in most undergraduate physical chemistry texts.
\bibitem{ie} In other words, one regards the exact rate as an approximation to the TST rate, instead of the other way round (since the `exact' rate describes a fictitious classical rate-process): see ref.\ \onlinecite{jor}.
\bibitem{WM} Wigner expanded the exponential to second order, from which he calculated the rate of transmission through a parabolic barrier.\cite{wiggie}
\bibitem{diff} This was difficult to compute numerically, since one must use a finite basis (in this case a Colbert-Miller DVR\cite{colbert}) to represent the density matrix and this is equivalent to approximating the Dirac-delta function in the flux operator by a finite-width function, which then gives a numerical approximation to \eqn{eq:cfsnf} that tends smoothly to zero as \shortta. We got around this by using a large basis (a total of 32768 grid-points between $\pm 150$ a.u.), which gave the correct behaviour down to about $t=1$ fs. We then matched to the \shortt value, which was computed by evaluating directly \eqn{eq:kwm}. The same technique was used to obtain the plots in Figs.~3b and 3c.
\bibitem{fourth} A further general point is that \eqn{eq:kwm} contains an $h(p)\equiv h({\hat p})$ term which can be regarded as a short time approximation to the projection operator onto final product states employed in Miller's earlier formulation of quantum rate theory \cite{mill}, in which the flux-side time-correlation function contains $h({\hat p})$ in place of the side operator, and has  a non-zero \shortt limit. This limit gives similar results to $k^\ddag_{\rm WM}(\beta)$, and suffers from the same statistical problems (discussed in Sec.~IIIC).
\bibitem{liudid} The emergence of bimodal structure in the Boltzmann matrix  has also been discussed by Liu and Miller in ref.\ \onlinecite{scivr3}.
\bibitem{inst} The complete instanton orbit is also a stationary point on the ring-polymer surface, but does not dominate the integral because it is a saddle point,  which the imposition of the $\delta(q)$-constraint 
does not turn into a minimum (because of the presence of a zero-frequency mode: see ref.~\onlinecite{equiv}).
\bibitem{paraprob} Note that also the \longt limit of \eqn{eq:cfsnf} is problematical for a parabolic barrier below $2T_{\rm c}$, since no amount of time $t$ is sufficient for $C_{\rm fs}^{[1]}(t)$ to recover from the initial quantum statistics in which the polymer stretches indefinitely on either side of the barrier. Interestingly, this problem is shared by \eqn{eq:cfsmil}, which means that, until \eqn{eq:cfsnfn} of this work (see Sec.~\ref{ssec:QTST}), there was no quantum flux-side time-correlation function capable of giving the correct parabolic barrier rate throughout the entire $T>T_{\rm c}$ temperature range across which it is finite.
\bibitem{shortcut} A much quicker way to derive \eqn{eq:knfn} is first to take the \shortt limit of $C_{\rm ss}^{[N]}(t)$, and then 
  to differentiate the result with respect to $t$. We have followed the longer derivation in order to establish the form of $C_{\rm fs}^{[N]}(t)$.
\end{thebibliography}
\end{document}